\documentclass[12pt,letterpaper,english]{article}

 \usepackage{setspace,graphics}
 \usepackage[dvips]{epsfig} 
 \usepackage{a4,amssymb,epsfig,array,cite}
 \usepackage[hypertex]{hyperref}
\usepackage{slashed}

\usepackage{amsfonts,bm,amssymb,euscript,array,babel}

\newcounter{multieqs}

\newtheorem{theorem}{Theorem}
\newtheorem{lemma}[theorem]{Lemma}

\newenvironment{proof}[1][Proof]{\begin{trivlist}
\item[\hskip \labelsep {\bfseries #1}]}{\end{trivlist}}

\newcommand{\qed}{\nobreak \ifvmode \relax \else
      \ifdim\lastskip<1.5em \hskip-\lastskip
      \hskip1.5em plus0em minus0.5em \fi \nobreak
      \vrule height0.75em width0.5em depth0.25em\fi}

\newcommand{\be}{\begin{equation}}
\newcommand{\ee}{\end{equation}}
\newcommand{\eq}[1]{(\ref{#1})}

\def\nn{\nonumber}
\def\bea{\begin{eqnarray}}
\def\eea{\end{eqnarray}}
\def\obar{\overline}

\def\beqa{\begin{eqnarray}} 
\def\eeqa{\end{eqnarray}} 
\def\beq{\begin{equation}} 
\def\eeq{\end{equation}}

\def\Tr{{\rm Tr}}

\def\a{\alpha}          
           
  \def\C{\Gamma}  
\def\d{\delta}

\def\g{\gamma}

\def\l{\lambda} \def\L{\Lambda} \def\la{\lambda}
\def\m{\mu}

\def\s{\sigma}

\def\cA{{\cal A}}  \def\cC{{\cal C}}
  
\def\cG{{\cal G}} \def\cH{{\cal H}} \def\cI{{\cal I}}
\def\cJ{{\cal J}}  \def\cL{{\cal L}}
\def\cM{{\cal M}} \def\cN{{\cal N}} \def\cO{{\cal O}}
  \def\cR{{\cal R}}
  \def\cU{{\cal U}}

\def\R{{\mathbb R}}
\def\C{{\mathbb C}}

\def\one{\mbox{1 \kern-.59em {\rm l}}}

\def\msu{\mathfrak{s}\mathfrak{u}}
\def\mmu{\mathfrak{u}}

\def\bit{\begin{itemize}}
\def\eit{\end{itemize}}

\def\({\left(}
\def\){\right)}
\def\diag{\mbox{diag}}

\def\Mat{{\rm Mat}}

\def\d{\delta}

 \def\del{\partial}

\def\uno{\mbox{1 \kern-.59em {\rm l}}}

\newcommand{\tr}{\mbox{tr}}

\def\Box{\square}

\def\bcomment#1{}

\renewcommand{\title}[1]{\vspace{10mm}\noindent{\Large{\bf #1}}\vspace{8mm}}
\newcommand{\authors}[1]{\noindent{\large #1}\vspace{5mm}}
\newcommand{\address}[1]{{\itshape #1\vspace{2mm}}}

\begin{document}


\begin{flushright}
UWTHPh-2010-4\\
\end{flushright}

\begin{center}

\title{Emergent Geometry and Gravity from Matrix Models:\\[1ex] 
an Introduction}

\authors{Harold {\sc Steinacker}${}^{1}$}

\address{Fakult\"at f\"ur Physik, Universit\"at Wien\\
Boltzmanngasse 5, A-1090 Wien, Austria}

\footnotetext[1]{harold.steinacker@univie.ac.at}

\vskip 1.5cm

\textbf{Abstract}

\vskip 3mm 

\begin{minipage}{14cm}%

A introductory review to emergent noncommutative gravity 
within Yang-Mills Matrix models is presented. Space-time is described as 
a noncommutative brane solution of the matrix model,
i.e. as submanifold of $\R^D$. 
Fields and matter on the brane arise as fluctuations of the 
bosonic resp. fermionic matrices around such a background, 
and couple to an effective metric interpreted in terms of gravity. 
Suitable tools are provided for the 
description of the effective geometry in the semi-classical limit. 
The relation to noncommutative gauge theory and the role of UV/IR mixing is explained. 
Several types of geometries are identified, in particular
"harmonic" and "Einstein" type of solutions. 
The physics of the harmonic branch is discussed in some detail, emphasizing the 
non-standard role of vacuum energy. This may provide new approach to some of the
big puzzles in this context. The IKKT model with $D=10$ 
and close relatives are singled out as promising
candidates for a quantum theory of fundamental interactions including gravity.

\end{minipage}

\end{center}


\setcounter{page}0
\thispagestyle{empty}
\newpage

\begin{spacing}{.3}
{
\noindent\rule\textwidth{.1pt}            
   \tableofcontents
\vspace{.6cm}
\noindent\rule\textwidth{.1pt}
}
\end{spacing}


\section{Introduction}

This article is intended as a pedagogical and mostly self-contained
introduction to emergent geometry 
and gravity within Yang-Mills matrix models. 
The aim of this line of research is to describe the physics and the geometry  
of general noncommutative (NC) spaces in these matrix-models, 
and to assess their viability as a quantum theory of
space-time including matter, gauge fields and gravity.
We hope that the present review provides a useful
basis for further research in this direction.

Our starting point is the identification of 
a gravity sector within noncommutative gauge theory. NC gauge theory
has been considered previously as 
a deformation of Yang-Mills gauge theory, living on NC space. 
From that point of view,
it is well-known that the $U(1)$ sector of
$U(n)$ gauge theory on the Moyal-Weyl quantum plane $\R^n_\theta$ 
(which is the simplest example of a NC space) plays a special role:
it does not decouple from the remaining $SU(n)$ degrees of freedom, and 
its quantum effective action is drastically different from its commutative 
counterpart due to UV/IR mixing. These and other "strange" features have been  viewed as obstacles 
for the physical application of NC gauge theory, and a relation to gravity 
has been widely conjectured. This conjecture is 
corroborated further through string-theoretical matrix models such as the 
IKKT model \cite{Ishibashi:1996xs},
the description of gravitons in the BFSS matrix model and related models
\cite{Banks:1996vh,Kabat:1997sa,Ishibashi:2000hh,Taylor:2001vb,Kitazawa:2006pj}, 
and matrix-models describing curved or "fuzzy" 
quantum  spaces 
\cite{Alekseev:2000fd,Grosse:2004wm,Behr:2005wp,Azuma:2005pm,DelgadilloBlando:2008vi,Azuma:2004qe}.

More recently, Rivelles observed \cite{Rivelles:2002ez} that the coupling 
of matter to
NC $U(1)$ gauge fields can be rewritten in terms of an effective metric, thus 
pointing out a direct relation between NC gauge theory with geometry and gravity.
Such a relation was also advocated in 
\cite{Yang:2004vd,Yang:2006mn}, in particular
a relation between self-dual gauge fields and gravitational instantons. 
This motivated a series of papers 
\cite{Steinacker:2007dq,Steinacker:2008ri,Grosse:2008xr,
Klammer:2008df,Steinacker:2008ya,Steinacker:2009mp} 
studying the effective geometry of general NC branes of Yang-Mills 
matrix models,
which provide a non-perturbative formulation of NC gauge theories.
The central point in these works
is that NC gauge theory is in fact a theory of noncommutative 
spaces and their fluctuations. The geometry and the 
effective metric is not a fundamental degree of freedom which is put in 
by hand, rather it arises effectively in the low-energy or semi-classical 
description of the model. That is the reason for
calling it "emergent NC gravity".

In order to have a well-defined framework, we will focus on matrix models
of Yang-Mills type. These models have
non-commutative spaces or space-time as solutions, 
i.e. quantized Poisson manifold.
Thus space-time and geometry are dynamical rather than put in by hand,
and the models should be considered as background independent.
$U(1)$ fluctuations of the matrices around  NC space-time describe
geometrical deformations such as gravitons, while $SU(n)$ fluctuations 
describe nonabelian gauge fields. The kinetic terms of these fields 
arises from the commutators in the matrix model, and 
encodes an effective metric which is essentially universal for 
all fields and matter. Since this metric is dynamical, it 
must be interpreted in terms of gravity.
This leads to an intrinsically non-commutative mechanism for gravity,
combining the metric and the Poisson 
structure in a specific way. It provides a natural
role for non-commutative or quantized space-time in physics.

Let us describe some results in more detail.
Space-time is described as a 3+1-dimensional 
NC brane $\cM_\theta\subset \R^{D}$ 
(possibly with compactified extra dimensions),
which carries a Poisson tensor $\theta^{\mu\nu}(x)$.
All matter and gauge fields live on this 
space-time brane, and there are no physical fields
propagating in the ambient $D$-dimensional space unlike in
other braneworld scenarios such as \cite{ArkaniHamed:1998rs}.
An effective metric 
$G^{\mu\nu}\sim \theta^{\mu\mu'}\theta^{\nu\nu'} g_{\mu'\nu'}$ arises on this 
space-time brane, which governs the kinetic term of
all fields more-or-less as in general relativity (GR). 
This metric is dynamical, however it is not a fundamental degree of 
freedom: it is determined by the embedding
$\cM_\theta \subset \R^{D}$, and the Poisson tensor
$\theta^{\mu\nu}$ describing noncommutativity. 
Hence the fundamental degrees of freedom are different from GR,
and can be interpreted alternatively in terms of NC gauge theory.
This makes the dynamics of emergent NC gravity 
somewhat difficult to disentangle, and 
the effective metric is not governed in general by the Einstein equations.

We will identify 2 classes of solutions: in the 
"Einstein branch", solutions of the Einstein equations can be 
realized as embedded submanifolds for $D \geq 10$. 
Since the Einstein-Hilbert action arises upon quantization and
is not part of the bare matrix model, 
the model must be free of UV/IR mixing above a scale
$\Lambda$ identified as Planck scale. This
singles out the IKKT model \cite{Ishibashi:1996xs} or close relatives, 
with $D=10$ and 
maximal supersymmetry above $\Lambda$. 
In contrast, the solutions in the "harmonic branch"  are
governed by the brane tension rather than the induced Einstein-Hilbert term.
That branch is interesting because the physics of vacuum energy is 
different from GR, providing a new perspective 
notably for the cosmological constant. For example,
a cosmological solution is discussed which requires far 
less fine-tuning than the $\Lambda$CDM concordance model, featuring a 
big bounce. Newtonian gravity arises in this branch 
through an interesting mechanism, with long-distance modifications 
which may be relevant for the issue of dark matter.
Even though the solar system precision tests appear to be violated
in the present ansatz, it seems conceivable that some of these features
carry over into a modified Einstein branch.

Apart from possible deviations from GR and its intrinsic appeal, 
the main motivation for the matrix model approach
is that notably the IKKT model promises to provide an accessible 
{\em quantum} theory of
gravity as well as the other fundamental interactions. The reason is 
that it can be equivalently interpreted as $\cN=4$ NC SYM on $\R^4_\theta$.
Thus it is (expected to be) UV finite, 
and free of pathological UV/IR mixing \cite{Jack:2001cr,Matusis:2000jf}. The 
remaining mild
UV/IR mixing below the $\cN=4$ SUSY breaking scale is 
nothing but induced gravity and therefore welcome. 
In particular, the quantization amounts to the quantization
of non-commutative Yang-Mills action, 
rather than the Einstein-Hilbert action. This 
elegantly circumvents many of the 
technical problems in quantizing GR. 
Hence these models under consideration 
define {\em some} accessible quantum theory of gravity, which is at least 
not too far from real gravity. 

Studying the geometrical properties of the matrix models, 
we will obtain structures and features which are familiar from string
theory. For example, the effective metric on the brane is
very similar to the open-string metric on branes in analogous string theory
considerations. This is to be expected in view of the relation with string
theory. However, the matrix model is taken as
a starting point here, because it is intrinsically non-perturbative
and extremely simple. Hence we cannot just use results
from string theory, rather everything will be derived in a
self-contained manner starting from the matrix model. 


Some words on the ``philosophy'' of emergent NC gravity 
are in order here\footnote{For a review of 
analogue models for emergent gravity see \cite{Barcelo:2005fc}.}.
Rather than trying to carry over some formulation of 
GR into the context of 
quantized spaces\footnote{which would not simplify
the problem of quantizing gravity in any obvious way.},
one starts with a model whose fundamental degrees of freedom 
are different from GR, but which provides a
physically viable description of geometry and gravity 
in some geometrical phase or limit.
Matrix models are not only very natural in the context of quantum spaces, 
they appear to realize the idea of emergent gravity, 
and moreover - most importantly - promise to define a good quantum theory.

In order to provide a coherent picture, we 
focus on the matrix model approach following
\cite{Steinacker:2007dq,Grosse:2008xr,Klammer:2008df,Steinacker:2008ri}
and provide the necessary background, rather than 
attempting a survey of related work and other points of view.
In particular, the relation with string theory 
is discussed only briefly in section \ref{sec:model}. 
This allows to keep the presentation simple 
and accessible, but does not imply
any disregard of other results which are not covered. 
The reader is assumed to have a basic background in 
high-energy theory, gauge theory and general relativity, but no 
previous knowledge in noncommutative field theory is required.

The structure of this review is as follows: 
in sections 2 and 3 the basic geometrical
properties of noncommutative branes are derived. In section 4 
a special class of geometries is discussed which appears to be most 
relevant for physical applications. The quantization 
(focusing on one-loop) is discussed in section 5.
The equations of motion for gravity 
coupled to matter are then discussed in section 6.
Section 7 is more speculative in nature, exploring the physics of 
the "harmonic branch" of solutions which is somewhat different from 
GR; section 7.4 contains some new material. 
The exploration of the more conventional "Einstein branch" and possible
other scenarios is left for future work.

\section{Yang-Mills matrix models and quantum spaces}
\label{sec:basic}

We consider the following type of Yang-Mills matrix models 
\be
S = - \frac{\L_0^4}{4 g^2}\,\Tr \Big([X^a,X^b] [X^{a'},X^{b'}] 
\eta_{aa'}\eta_{bb'} \Big) 
\label{YM-action-1}
\ee
where 
\be
\eta_{aa'} = \diag(-1,1,...,1);
\ee
the Euclidean version of the model is obtained 
by replacing 
\be
\eta_{aa'} \quad \to \quad \delta_{aa'} .
\ee
The degrees of freedom of this model are 
hermitian\footnote{in the Minkowski case, 
we will assume that the time-like matrices are anti-hermitian. 
This will be addressed in more detail below.} 
matrices $X^a\,\in \Mat(\infty,\C)$ for $a=0,1,2,..., D-1$.
We introduced also an energy scale $\Lambda_0$ 
which gives the matrices $X^a$ the
dimension of length.
The action is invariant under the fundamental gauge symmetry
\be
X^\mu \to U X^\mu U^{-1}, \qquad U \in \cU(\infty) 
\label{gauge}
\ee
as well as a global rotational and translational
symmetry, where rotations act as $X^a \to \Lambda^a_b\, X^b$ 
for $\Lambda^a_b \in SO(D-1,1)$ resp. $SO(D)$, and 
translations $X^a \to X^a + c^a \one$ for $c^a \in \R$. 
However there is no space-time 
or geometry to start with; space and geometry only emerge
on a given solution of the model, and are therefore  dynamical. 
The only geometrical input is the constant metric 
$\eta_{ab}$ resp. $\d_{ab}$ for the ``embedding space'' $\R^D$.
This space is however unphysical and only serves to ``suspend''
the physical space-time brane and its emergent geometry, 
as we will see. Fermionic matter will be added
in section \ref{sec:fermions}. 

This class of models can be obtained as dimensional reduction of
Yang-Mills gauge theory to a point. In particular,
the IKKT model with $D=10$ is singled out by an extended 
matrix supersymmetry \cite{Ishibashi:1996xs}. 
This is probably essential to obtain a 
well-defined quantum theory.
Some modifications of these models are conceivable, such as
a "mass" term $Tr(m^2 \eta_{ab} X^a X^b)$, or
cubic terms $Tr(Q_{abc} X^a [X^b, X^c])$. These
``soft'' terms will
have mild impact on the quantization properties on a 4-dimensional
space-time brane, and allow to obtain interesting 
compactification scenarios.

The equations of motion corresponding to the action \eq{YM-action-1} 
is obtained by varying the entries of the matrices $X^a$, which
gives
\be
[X^b,[X^{a},X^{b'}]]\, \eta_{b b'} = 0 \, .
\label{eom}
\ee
This equation has various types of solutions. An obvious solution 
is given by any commuting set of matrices $[X^a,X^b] = 0$.
These "commutative" configurations do not support propagating fields in this model
and are in some sense singular; we will not pursue them any further here. 
The prototype of the class of solutions which is important here is 
given by\footnote{The particular embedding of $\R^{2n} \subset \R^D$
implied by this choice of variables is of course arbitrary, and will be 
generalized below.}
\be
X^a \equiv \left(\begin{array}{c}X^\mu \\ \phi^i\end{array}\right) 
= \left(\begin{array}{c}\bar X^\mu \\ 0\end{array}\right), \qquad \mu = 0,...,2n-1, 
\quad i=1, ..., D-2n 
\label{extradim-splitting}
\ee
where $\bar X^\mu$ are generators of the Moyal-Weyl quantum plane
$\R^{2n}_\theta$,
\be
[\bar X^\mu,\bar X^\nu] = i\bar\theta^{\mu\nu}\, \one, 
     \qquad \mu, \nu= 0,...,2n-1 .
\label{Moyal-Weyl}
\ee
Here $\bar\theta^{\mu\nu}$ is a 
constant antisymmetric numerical tensor.
This solution should be interpreted as 
quantization of flat $\R^{2n} \subset \R^D$ with Poisson structure 
$\{x^\mu,x^\nu\} = \bar\theta^{\mu\nu}$. 
From a mathematical
point of view this is very similar to quantum mechanics, 
where phase space is quantized using the same commutation relations.
The matrices
$X^\mu$ should thus be interpreted as quantized coordinate functions 
on $\R^{2n}$. The matrices $\phi^i$ can similarly
be interpreted as quantized functions resp. scalar fields 
on $\R^{2n}$, which happen to vanish in this particular solution.
As in quantum mechanics,
the above commutation relations imply
space-time coordinate uncertainty relations 
\be
\Delta x^\mu \Delta x^\nu \geq \frac 12 \, |\bar\theta^{\mu\nu}| .
\label{uncertainty}
\ee
This means that space-time becomes ``fuzzy'' at
the noncommutative scale defined by 
\be
\Lambda_{NC}^{2n} = (\det\bar\theta^{ab})^{-1/2}  .
\label{Lambda-NC}
\ee
Such uncertainty relations can be
motivated by standard arguments combining quantum 
gravity with mechanics \cite{Doplicher:1994tu}.

The important message is that space-time arises
as a {\em solution} of the matrix model, 
it is {\em not} a fixed background.  This suggests some relation with gravity.

\subsection{Quantized Poisson manifolds.}
\label{sec:quant-poisson}

To proceed, we recall the concept of the quantization 
of a Poisson manifold $(\cM,\{.,.\})$, referring e.g. to 
\cite{Bordemann:1993zv} and references therein 
for more mathematical background. 
A Poisson structure is an anti-symmetric bracket 
$\{.,.\}:\,\cC(\cM)\times \cC(\cM) \to \cC(\cM)$ 
which is a derivation in each argument and satisfies the Jacobi identity,
\be
\{fg,h\} = f\{g,h\} + g\{f,h\}, \qquad \{f,\{g,h\}\} + {\rm cycl.} = 0 .
\ee
It is sometimes useful to introduce 
an expansion parameter of dimension ${length}^2$ and write
\be
\{x^\mu,x^\nu\} = \theta^{\mu\nu} = \theta\,\theta_0^{\mu\nu}(x)
\ee
where $\theta_0^{\mu\nu}(x)$ 
is some fixed Poisson structure.
Given a Poisson manifold, we denote as
{\em quantization map} an isomorphism of vector spaces
\be
\begin{array}{rcl}
\cI: \quad \cC(\cM) &\to& \cM_\theta\,\,\subset \,\, \Mat(\infty,\C)\, \\
 f(x) &\mapsto& F
\end{array}
\label{quant-map} 
\ee
which depends on the Poisson structure $\cI \equiv \cI_\theta$, 
and satisfies\footnote{The precise definition of this
limiting process is non-trivial, and there
are various definitions and approaches. 
Here we simply
assume that the limit and the expansion in $\theta$ exist
in some appropriate sense.}
\be
\cI(f g) - \cI(f)\cI(g) \,\, \to \,\, 0 \quad\mbox{and}\quad
\frac 1\theta \Big(\cI(i\{f,g\}) - [\cI(f),\cI(g)]\Big) \,\, \to \,\, 0 
\qquad \mbox{as}\quad \theta \to 0 .
\label{poisson-comp}
\ee
Here $\cC(\cM)$ denotes some space of functions on $\cM$, 
and $\cM_\theta$ is interpreted as quantized algebra of 
functions\footnote{$\cA$ is the algebra 
generated by $X^\mu = \cI(x^\mu)$, or some 
subalgebra corresponding to well-behaved functions.} 
 on $\cM$. Such a quantization map $\cI$ is not unique, 
i.e. the higher-order terms in \eq{poisson-comp} are not unique.
For example, the Moyal-Weyl
quantum plane $\R^{2n}_\theta$ \eq{Moyal-Weyl} 
is a quantization of $\R^{2n}$
with constant Poisson tensor $\bar\theta^{\mu\nu}$, and
a natural (``Weyl'') quantization map is given by 
\be
\begin{array}{rcl}
\cI: \quad \cC(\R^{2n}) &\to& \R^{2n}_\theta\,\,\subset \,\, \Mat(\infty,\C)\, \\[1ex]
 e^{i k_\mu x^\mu} &\mapsto& e^{i k_\mu \bar X^\mu} .
\end{array}
\label{MW-map} 
\ee
The map $\cI$ allows to define a 
``star'' product on $\cC(\cM)$ as the pull-back of the algebra resp. 
matrix product in $\cM_\theta$,
\be
f \star g := \cI^{-1}(\cI(f) \cI(g)) 
\label{star-product}
\ee 
which allows to work with classical functions, 
hiding $\theta$ in the star product.
In the example of $\R^{2n}_\theta$, it leads
to the well-known Groenewald-Moyal star product on $\R^{2n}$. 
The compatibility \eq{poisson-comp}
with the Poisson structure is encoded by the requirement 
\be
f \star g = f g + \frac 12 \{f,g\} + O(\theta^2) .
\ee
Kontsevich has shown \cite{Kontsevich:1997vb} that such a quantization 
always exists in the sense of formal power series in $\theta$.
This is a bit too weak for the present context since 
we deal with operator resp. matrix quantizations. However
we will assume that $\theta^{\mu\nu}$ is non-degenerate defining
a symplectic structure
\be
\omega = \frac 12 \theta^{-1}_{\mu \nu} dx^\mu\wedge dx^\nu,
\ee 
for which stronger mathematical results 
are available.

\paragraph{De-quantization and semi-classical limit.}

Since the matrix model provides directly quantum spaces
such as $\R^{2n}_\theta$,
we are actually faced with the opposite problem of ``de-quantization'', 
i.e. the semi-classical limit of a quantum space.
The above concepts can be used as a tool to understand solutions of
the matrix model in the language
of ordinary spaces and effective actions.
Denoting the inverse of the quantization map 
as $f_\theta := \cI_\theta^{-1}(F)$, we can replace
$[F,G]$ by $[f_\theta,g_\theta]_\star$ with $\cI$ being understood.
Now we can define the "semi-classical limit" as the 
leading non-vanishing term in an expansion in $\theta$,
dropping all higher-order terms in $\theta$;
this semi-classical limit will be indicated by $\sim$. 
In particular, 
\be
[F,G] \,\,\sim\,\, i \{f,g\} \qquad\quad (+ O(\theta^2) )
\ee
which will be used throughout.
Thus one can simply replace commutators with Poisson brackets
in the semi-classical limit, and any ambiguity in $\cI_\theta$ would
show up only in higher-order corrections; this is familiar from quantum
mechanics.
In particular, if some generators $X^\mu$ generate the entire operator
resp. matrix algebra, we can at least formally write every
``smooth function'' on $\cM_\theta$ as $F = F(X^\mu)$.
Then 
\be
X^\mu \sim x^\mu
\ee
can be interpreted as quantized coordinate function on $\cM$, and
we can write 
\be
[X^\mu,F] \sim i\theta^{\mu\nu}(x) 
\frac{\partial}{\partial x^\nu} f(x) .
\label{derivation}
\ee
The important point is that matrices $F \in \Mat(\infty,\C)$
should be interpreted as quantizations of functions on $\cM$,
and their non-commutative product encodes an energy scale 
$\L_{NC}$. It is also helpful to recall the concept of
coherent states or optimally localized states, which 
saturate the uncertainty relations \eq{uncertainty}.
They make explicit the fact that on length scales larger than
$L_{NC} = \L_{NC}^{-1}$, the quantized spaces $\cM_\theta$
look and behave like their (semi-) classical limit $\cM$.
In much of this paper we will restrict ourselves
to the semi-classical limit of the matrix model 
and its geometrical solutions.

Finally, under favorable assumptions the integral on 
$\cM$ is related to the trace as follows (cf. \cite{Bordemann:1993zv})
\bea
\int \frac{\omega^n}{n!} f &=& 
\int \frac{d^{2n}x}{\theta(x)^{n}}\, f(x) \,\,\sim \,\,(2\pi)^n \Tr\, \cI(f) ,  
\label{volume-trace}\\
\theta(x) &:=& {\rm Pfaff}(\theta^{\mu\nu}(x))^{\frac 1n}
= |\theta^{\mu\nu}(x)|^{\frac 1{2n}} \,\,=: \,\, L_{NC}^{2}(x) .
\eea
The symplectic volume form is singled out by the compatibility 
condition
\be
\Tr[\cI(f),\cI(g)] \sim \int \frac{\omega^n}{n!} \{f,g\} = 0 .
\ee
It is instructive to recall here the Darboux theorem, which 
states that every symplectic 2-form 
is locally constant in suitable coordinates.
This means that in suitable coordinates, 
any quantum space (with non-degenerate $\theta^{\mu\nu}$) 
locally looks like $\R^{2n}_\theta$. It also
provides an intuitive justification for \eq{volume-trace},
which then reduces to the Bohr-Sommerfeld quantization rule.

\subsection{Moyal-Weyl solution and noncommutative gauge theory I}
\label{sec:moyal-weyl-gauge}

Let us start with the Moyal-Weyl solution \eq{Moyal-Weyl}, and recall 
that translations on $\R^{2n}_\theta$ are realized as 
inner automorphisms, 
\be
U f(\bar X^\nu) U^{-1} = f(\bar X^\nu - \bar\theta^{\mu\nu} k_\mu),
\qquad U = e^{i k_\mu \bar X^\mu} .
\label{translations-MW}
\ee
This justifies to define a partial derivative operator on
 $\R^{2n}_\theta$ in terms of a commutator
\be
\frac{\partial}{\del \bar X^\mu} f
= -i\bar\theta^{-1}_{\mu\nu} [\bar X^\nu,f] ,
\label{partial-MW}
\ee
consistent\footnote{the Poisson structure 
is always assumed to be non-degenerate.} with the semi-classical limit \eq{derivation}.
The fact that derivatives are inner derivations
constitutes the basic difference between the commutative and
the non-commutative framework.
This is compatible with the definition of the integral over $\R^4_\theta$ as
\be
\int_{\R^4_\theta} f := \frac{(2\pi)^2}{\bar\theta^{2}} \Tr f ,
\qquad \bar\theta^{2} = \sqrt{|\bar\theta^{\mu\nu}(x)|} \,= \, \L_{NC}^{-4} 
\ee
which has the correct classical limit
\eq{volume-trace}, and satisfies $\int_{\R^4_\theta} \del_\mu f = 0$
 up to ``boundary terms''.

Now consider fluctuations of the above solution 
$\R^{4}_\theta\subset \R^D$ \eq{extradim-splitting},
\eq{Moyal-Weyl} of the matrix model, 
focusing on the 
4-dimensional case to be specific.
Such fluctuations can be parametrized as 
\be
\left(\begin{array}{c}X^\mu \\ \phi^i\end{array}\right) 
= \left(\begin{array}{c}\bar X^\mu \\ 0\end{array}\right)
 + \left(\begin{array}{c}\cA^\mu \\ \phi^i\end{array}\right)
\label{cov-coord-1}
\ee
where $\cA^\mu$ and $\phi^i$ are ``small'' hermitian matrices.
As explained above, they can be  interpreted
as smooth functions on $\R^4_{\theta}$, i.e. 
$\cA^\mu =\cA^\mu(\bar X) \sim \cA^\mu(x)$ and 
$\phi^i = \phi^i(\bar X) \sim \phi^i(x)$.
Consider now the
change of variables
\be
\cA^\mu = -\bar\theta^{\mu\nu} A_\nu
\label{A-naive}
\ee
where $A_\nu$ is hermitian. 
Using \eq{partial-MW} one finds
\bea
[X^\mu,f] &=& [\bar X^\mu + \cA^\mu,f] = i \bar\theta^{\mu\nu} 
(\frac{\partial}{\partial \bar X^\nu} f
+ i [A_\nu,f]) \,\,\equiv\,\,  i \bar\theta^{\mu\nu} D_\nu f, \nn\\
\,[X^\mu,X^\nu] 
&=& -i \bar\theta^{\mu\mu'} \bar\theta^{\nu\nu'}
(\bar\theta^{-1}_{\mu'\nu'} + F_{\mu'\nu'})
\label{XX-gauge}
\eea 
where $F_{\mu\nu} = \partial_\mu A_\nu - \partial_\mu A_\nu + i [A_\mu,A_\nu]\,$ is the
$U(1)$ field strength on $\R^4_\theta$.
The symmetry \eq{gauge} of the matrix model now acts 
on the fluctuation fields as 
\be
A_\mu \to U A_\mu U^{-1} + U \partial_\mu U^{-1}, \qquad
\phi^i \to U \phi^i U^{-1} .
\label{gauge-A-phi}
\ee
This clearly has the structure of gauge transformations 
of Yang-Mills gauge fields\footnote{The 
$X^\mu$ can be interpreted 
as ``covariant coordinates'' \cite{Madore:2000en}.}. 
Moreover, the action \eq{YM-action-1} can now be rewritten 
as 
\bea
S &=& S[A] + S[\phi]  
 = \frac{1}{(2\pi)^2} \int_{\R^4_\theta} d^4 x\,
 \frac{\Lambda_0^4\bar\theta^{2}}{g^2} \, 
\Big(\bar G^{\mu\mu'}\,\bar G^{\nu\nu'}\,F_{\mu\nu}\,F_{\mu'\nu'}
\, + \bar G^{\mu\nu} \eta_{\mu\nu}  \nn\\
&& \qquad 
 +  \frac 2{\bar\theta^{2}}\, \bar G^{\mu\nu}\, D_\mu\phi^i D_\nu \phi^j \d_{ij}
 + \frac 1{\bar\theta^{4}} \,
 [\phi^i,\phi^j][\phi^{i'},\phi^{j'}] \d_{ii'} \d_{jj'}\Big) 
\label{scalar-action-A-tilde}
\eea
dropping surface terms, where 
\be
\bar G^{\mu\nu} = \frac 1{\bar\theta^{2}}\,\bar\theta^{\mu\mu'} \bar\theta^{\nu\nu'} 
\eta_{\mu'\nu'} , \qquad |\bar G^{\mu\nu}| = 1 .
\label{metric-MW}
\ee
These formulas are exact (up to boundary terms) and define some 
noncommutative field theory \cite{Douglas:2001ba}.
Thus the matrix model \eq{YM-action-1} on the solution
$\R^4_\theta$ can be viewed as a 
noncommutative gauge theory coupled to scalar fields $\phi^i$ 
in the adjoint, with effective flat metric $\bar G^{\mu\nu}$. 
Redefining $\tilde\phi^i = \bar\theta^{-1}\,\phi^i$
gives the scalar fields the usual dimension of mass.

\paragraph{Nonabelian gauge fields.}

Nonabelian gauge fields arise very naturally in 
the matrix model. Observe first that
for every solution $X^a$ of \eq{eom},
\be
X^a \otimes \one_n
\ee
is also a solution, naturally interpreted as 
$n$ coincident branes. The fluctuations around such a
background can be parametrized as
\be
\left(\begin{array}{c}X^\mu \\ \phi^i\end{array}\right) 
= \left(\begin{array}{c}\bar X^\mu \otimes \one_n \\ 0\end{array}\right)
 + \left(\begin{array}{c}\cA^\mu \\ \phi^i \end{array}\right),
\label{cov-coord-1-new}
\ee
considering $\R^4_\theta$ to be specific.
In analogy to the above considerations,
it is easy to see that
\be
\cA^\mu = -\theta^{\mu\nu} A_{\nu,\a}(\bar X) \l^\a, \qquad
\phi^i = \phi^i_\a(\bar X) \l^a
\ee
describe $\mmu(n)$- valued gauge resp. scalar fields on 
$\R^4_\theta$, denoting with $\l^\a$ a basis of $\mmu(n)$.
The matrix model action for the fluctuations then takes the form
\bea
S &=& \frac{1}{(2\pi)^2} \int_{\R^4_\theta} d^4 x\,
\frac{\Lambda_0^4\bar\theta^{2}}{g^2}\, 
\tr \Big(\bar G^{\mu\mu'}\,\bar G^{\nu\nu'}\,F_{\mu\nu}\,F_{\mu'\nu'}
\, + \bar G^{\mu\nu} \eta_{\mu\nu}\one_n  \nn\\
&& \qquad 
 +  \frac 2{\bar\theta^{2}}\, \bar G^{\mu\nu}\, D_\mu\phi^i D_\nu \phi^j \d_{ij}
 + \frac 1{\bar\theta^{4}} \,
 [\phi^i,\phi^j][\phi^{i'},\phi^{j'}] \d_{ii'} \d_{jj'}\Big) 
\eea
where $\tr()$ denotes the trace over the $\mmu(n)$ matrices,
and $F_{\mu\nu}$ is the $\mmu(n)$ field strength.

This mechanism is very remarkable: 
It shows that gauge theory arises ``automatically'' in the
matrix model, there is no need to define sophisticated 
mathematical structures such as principal fiber bundles and
connections.
This is one of the reasons why these models
are so interesting. On the other hand, 
even though the derivation and interpretation in terms 
of gauge theory on $\R^4_\theta$
seems impeccable, 
it is nevertheless misleading 
and physically ``wrong'' for the $U(1)$ sector, 
for a number or reasons:

\begin{itemize}

\item
While $U(n)$ gauge fields arise naturally on 
$\R^4_\theta$, there seems to be no way to 
separate them into $U(1)$ and $SU(n)$ gauge fields.
This is fundamentally 
tied to the noncommutativity of space. 
For example, if we were to impose 
the constraint that $A_{\mu}$
is traceless $\tr A_{\mu}=0$, then a
gauge transformation 
$\d A_\mu = \partial_\mu \Lambda + [A_\mu,\Lambda] $ 
will re-introduce a trace-$U(1)$ component $\sim\one_n$. 
Indeed the trace-$U(1)$ components 
of gauge fields and scalar fields 
are in-separably entangled with all other fields. We will see later
that this is nothing but a gravitational coupling, 
and the $U(1)$ components will be understood as geometrical
degrees of freedom. Only the $SU(n)$ components of the gauge field
turn out to be physical gauge fields, while the 
trace-$U(1)$ components turn into gravitational waves.
The latter will be denoted as ``would-be $U(1)$ gauge fields''
henceforth.

\item
We will see in section \ref{sec:quantization} that upon quantization, 
the one-loop effective action 
contains ``strange'' UV/IR mixing terms for the
trace-$U(1)$ component, but not for the $SU(n)$ terms. 
This indicates that the would-be $U(1)$ gauge fields
should not be interpreted as photons.
These ``strange'' terms will be understood 
as induced gravitational terms.
A notable exception is the $N=4$ SUSY case, which contains no
such terms \cite{Matusis:2000jf}.

\item 
Eq. \eq{translations-MW} says that
translations on $\R^4_\theta$
are nothing but particular gauge transformations. 
Thus the $U(1)$ gauge transformations on noncommutative
space contain space-time transformations, which 
can be interpreted as symplectomorphisms which form
a subgroup of the volume-preserving diffeomorphisms.
In the same vein, there are no  local observables in 
noncommutative gauge theory: for example, 
the $U(1)$ field strength $F_{\mu\nu}$ is not 
invariant under $U(1)$, but transforms as a 
scalar field under these symplectomorphisms. 
Gauge invariant observables such as
$\Tr (e^{i k_\mu X^\mu})$ 
involve an integral over space. 
All these are signatures of gravity.

\item
Slightly modified versions of the matrix model 
have quantum spaces with non-trivial geometries 
such as fuzzy spheres as solutions, cf. 
\cite{Madore:1991bw,Alekseev:2000fd,Iso:2001mg,Grosse:2004wm}. 
It is then obvious that fluctuations of these
solutions, i.e. fluctuations of the matrices,
should correspond to fluctuations of the geometry
rather than $U(1)$ gauge resp. scalar fields.

\end{itemize}

The resolution of this puzzle is the key for identifying
emergent gravity in the matrix model.
In the geometrical
interpretation explained below, the trace- $U(1)$ components
of $A_\mu(x)$ and $\phi^i$ will completely
absorbed in the effective metric $G^{\mu\nu}(x)$,
leaving only physical $SU(n)$ gauge fields and scalars 
as well as fermionic matter coupled to gravity.

\subsubsection{Euclidean versus Minkowski signature.}
\label{sec:euclid-mink}

Let us take a closer look at the effective 
metric $\bar G^{\mu\nu}$ for gauge fields on 
$\R^4_\theta$, focusing first on the Euclidean case.
Using a $SO(4)$ rotation, we can assume that
$\theta^{\mu\nu}$ has the canonical form
\be
\theta^{\mu\nu} = \theta\,\left(\begin{array}{cccc} 0 & 0 & 0 & -\a \\
                                0 & 0 & \pm\a^{-1} & 0 \\
                                0 & \mp\a^{-1} & 0 & 0  \\
                                \a & 0 & 0 & 0 \end{array}\right)\, .
\label{theta-standard-general-E}
\ee 
Clearly the corresponding symplectic 2-form
$\omega = \theta^{-1}_{\mu \nu} dx^\mu\wedge dx^\nu$
is (anti-) self-dual
$\star \omega = \pm \omega$  if and only if 
$\a^2 = 1$, where $\star$ denotes the Hodge star
defined by $\varepsilon^{\mu\nu\rho\sigma}$ and 
$\d_{\mu\nu}$ on $\R^4$.
On the other hand, \eq{metric-MW} gives 
\be
\bar G^{\mu\nu} = \frac 1{\theta^{2}}\,\theta^{\mu\mu'} \theta^{\nu\nu'} 
\d_{\mu'\nu'} 
= \diag(\a^2,\a^{-2},\a^{-2},\a^2) .
\label{G-MW-E-1}
\ee
Therefore $\bar G^{\mu\nu} = \d^{\mu\nu}$ if and only if
$\omega$ is (anti-)self-dual, otherwise $\bar G^{\mu\nu}$
differs from the embedding metric $\d^{\mu\nu}$
via a volume-preserving linear transformation.

Now consider the physical case of Minkowski signature
with embedding metric $\eta_{\mu\nu} = (-1,1,1,1)$.
The standard point of view in NC field theory is that
$\theta^{\mu\nu}$ is real-valued. Using suitable 
Lorentz transformations, we can again assume that it
has the above form \eq{theta-standard-general-E}, so that
\be
\bar G^{\mu\nu} = \frac 1{\theta^{2}}\,\theta^{\mu\mu'} \theta^{\nu\nu'} 
\eta_{\mu'\nu'} 
= \diag(\a^2,\a^{-2},\a^{-2},-\a^2) .
\ee
This indeed has Minkowski signature, but the role of 
time has switched from the first to the 
last coordinate. In particular there is 
no way that the effective metric $\bar G^{\mu\nu}$
agrees with the embedding metric $\eta^{\mu\nu}$.
This is a priori not a problem since only 
$\bar G^{\mu\nu}$ couples to the physical fields. 
However it is quite counterintuitive, and becomes
more problematic at the one-loop level as we will see.
To gain some insight
consider again the Euclidean case   and write
\be
[X^\mu,X^\nu] = i \theta^{\mu\nu}_E
\ee
with real $\theta^{\mu\nu}_E$. 
Having in mind a Wick rotation of the matrices 
\be
X^0 \to i T
\ee
such that the coordinate functions 
are related as usual via $x^0 \to it$,
the above commutator becomes
\be
\, [T,X^i] = \theta^{0 i}_E =: i \theta^{0 i}_M, \qquad
\,[X^i,X^j] = i \theta^{ij}_E =: i \theta^{ij}_M .
\ee
This leads to a generalized 
Poisson tensor $\theta^{\mu\nu}_M$ 
with imaginary time-like components, according to
$(X^\mu)^* = \eta_{\mu\nu} X^\nu$.
We can again assume that it has the form
\be
\theta^{\mu\nu}_M = \theta\,\left(\begin{array}{cccc} 0 & 0 & 0 & -i\a \\
                                0 & 0 & \pm\a^{-1} & 0 \\
                                0 & \mp\a^{-1} & 0 & 0  \\
                                i\a & 0 & 0 & 0 \end{array}\right)\, .
\label{theta-standard-general-M-imag}
\ee 
in suitable coordinates.
Now the effective metric \eq{metric-MW} has the form 
\be
\bar G^{\mu\nu} = \frac 1{\theta^{2}}\,\theta^{\mu\mu'}_M \theta^{\nu\nu'}_M 
\eta_{\mu'\nu'} 
= \diag(-\a^2,\a^{-2},\a^{-2},\a^2) ,
\label{G-MW-E}
\ee
and the time coordinate is in the expected slot.
Moreover,
$\star \omega = \pm \omega$ if and only if 
$\a^2 = 1$, where $\star$ denotes the Hodge star
defined by $\eta_{\mu\nu}$ on $\R^4$;
here we adopt the convention that the Wick-rotated epsilon-tensor is 
given by $\varepsilon^{0123} = i$.
Thus we again have
$\bar G^{\mu\nu} = \eta^{\mu\nu}$ if and only if 
$\omega$ is (anti-) self-dual $\star \omega = \pm \omega$, 
as in the Euclidean 
case.  This seems to be the appropriate concept of 
Wick rotation in the matrix model framework,
which will generalize to the case of non-trivial 
geometries and gravity.

\section{Geometry and gravity from matrix models}
\label{sec:model}

We have seen that NC space-time is obtained as a 
solution of the matrix model. Hence
space-time is dynamical, which strongly points at gravity. 
On the other hand, 
the fluctuations of space-time were identified as 
gauge fields on $\R^4_{\theta}$. 
The key observation of emergent NC gravity is that these
apparently different interpretations 
are two ways of looking at the same thing. 
This leads to an intrinsically 
non-commutative and compelling mechanism for 
gravity, realizing ideas in  \cite{Rivelles:2002ez,Yang:2004vd}. 
Here we follow a systematic approach and derive everything
from the framework of matrix models.

We consider configurations $X^a$ of the matrix model 
which can be interpreted as quantized embedding functions 
\be
X^a \sim x^a:\quad \cM^{2n}_\theta \hookrightarrow \R^{{D}}
\ee 
determining some 
$2n$-dimensional submanifold ("brane") of $\R^D$. 
Here $\sim$ denotes the {\em semi-classical}  
limit as discussed in 
section \ref{sec:quant-poisson}. 
The intuitive picture is that there 
are sufficiently localized (``coherent'') states such that
$\langle X^a\rangle$ spans the
manifold $\cM^{2n} \subset \R^{{D}}$.
We can then split (at least "locally")  the set of matrices 
as in \eq{extradim-splitting},
\be
X^a = \left(\begin{array}{c}X^\mu \\ \phi^i\end{array}\right),
 \qquad \mu = 0,...,2n-1, 
\quad i=1, ..., D-2n 
\label{extradim-splitting-general}
\ee
such that the $2n$ generators $X^\mu$ 
generate the full matrix algebra\footnote{
Since $Mat(\infty,\C)$ is irreducible, this implies 
that $\theta^{\mu\nu}$ as defined in \eq{theta-induced} 
is non-degenerate.} 
$Mat(\infty,\C)$, 
and therefore $\phi^i = \phi^i(X^\mu) \sim \phi^i(x^\mu)$.  
Since the commutators
$[X^\mu,X^\nu]$ 
always satisfy the Jacobi identity, they can be 
interpreted as quantizations of a
Poisson structure on $\cM$,
\be 
[X^\mu,X^\nu] =:i\theta^{\mu\nu}(X^\mu)\, \sim\, i\theta^{\mu\nu}(x), 
\quad  \mu,\nu = 1,..., 2n 
\label{theta-induced} 
\ee  
Thus 
$Mat(\infty,\C) \cong \cC_\theta(\cM)$ is  
interpreted as quantized algebra of functions on 
a manifold $\cM$ with Poisson structure $\theta^{\mu\nu}(x)$. 
In particular, the $X^\mu \sim x^\mu$ define 
quantized (local) coordinate functions on $\cM$, which we will denote as
"matrix coordinates".

The difference to the discussion in section 
\ref{sec:moyal-weyl-gauge} is that 
the ``would-be $U(1)$ gauge fields
$\cA^\mu$ are absorbed in $X^\mu$, 
avoiding the unphysical splitting 
of the Poisson tensor $\theta^{-1}_{\mu\nu}$ 
in \eq{XX-gauge}.

Note that all physical fields in this framework arise from
fluctuations of the matrices
around such a background (leading to nonabelian gauge fields
and scalars), 
and from the fermionic matrices $\Psi$. Since 
$Mat(\infty,\C)\cong \cC_\theta(\cM)$, 
it follows that they all
live only on the brane $\cM$, hence there is no 
physical higher-dimensional ``bulk''
which could carry any propagating degrees of 
freedom\footnote{
This does not exclude the existence of compactified physical
extra dimensions in the matrix model.}.
Moreover all fields propagate according to an 
effective metric, which we now identify.

\subsection{Emergent geometry.}
\label{sec:emergent-geom}

To understand the effective geometry of $\cM^{2n}$,
consider a test-particle on $\cM^{2n}$, 
modeled by a scalar field $\varphi$ for simplicity
(this could be e.g. an $su(k)$ component of $\phi^i$).
In order to preserve gauge invariance, 
the kinetic term must have the form
\bea
S[\varphi] &\equiv& - \Tr [X^a,\varphi][X^b,\varphi] \eta_{ab} 
\sim \frac{1}{(2\pi)^n}\, \int d^{2n} x\; \frac 1{\theta^{n}}\,\,
e^a(\varphi)  e^a(\varphi) \eta_{ab} \nn\\
&=&   \frac{1}{(2\pi)^n}\, \int d^{2n} x\; \frac 1{\theta^{n}}\,
 \theta^{\mu\mu'}(x) \theta^{\nu\nu'}(x) g_{\mu\nu}
\partial_{\mu'}\varphi \partial_{\nu'}\varphi  \nn\\
 &=&\frac{1}{(2\pi)^n}\, \int d^{2n} x\; 
\sqrt{|G_{\mu\nu}|}\,G^{\mu\nu}(x)
 \partial_{\mu} \varphi \partial_{\nu} \varphi \,,
\label{covariant-action-scalar}
\eea
denoting the $D$ natural vector fields on $\cM$ defined
by the matrix model as 
\be
e^a(f) := -i [X^a,f] \,\sim\, \theta^{\mu\nu}\partial_\mu x^a \partial_\nu f
\ee
where \cite{Steinacker:2008ri}
\bea  
G^{\mu\nu}(x) &:=& e^{-\sigma}\,\theta^{\mu\mu'}(x) \theta^{\nu\nu'}(x) 
 g_{\mu'\nu'}(x)  
\label{G-def-general}  \\
g_{\mu\nu}(x) &:=& \partial_\mu x^a \partial_\nu x^b \eta_{ab} \,\,
= \eta_{\mu\nu}(x)  + \partial_\mu \phi^i \partial_\nu \phi^j \d_{ij},
\label{g-explicit}\\
e^{-(n-1)\sigma} &:=& \frac 1{\theta^{n}}\, |g_{\mu\nu}(x)|^{-\frac 12},
\qquad \theta^n = |\theta^{\mu\nu}|^{1/2} .
\label{sigma-rho-relation} 
\eea
Here $g_{\mu\nu}(x)$ 
is the metric induced on $\cM\subset \R^{D}$ via 
pull-back of $\eta_{ab}$. The normalization factor 
$e^{-\sigma}$ is determined uniquely such that
\be
\frac 1{\theta^{n}} = \sqrt{|G_{\mu\nu}|}\, e^{-\sigma} ,
\label{rho-sigma-det}
\ee 
except for $n=1$ which we exclude for simplicity.
Therefore the kinetic term for $\varphi$ on  $\cM$
is governed by the metric $G_{\mu\nu}(x)$, 
which depends on the Poisson tensor $\theta^{\mu\nu}(x)$ and the 
embedding 
metric $g_{\mu\nu}(x)$. We will see below  that
the same metric also governs nonabelian gauge fields 
and fermions  
in the matrix model (up to possible conformal factors), 
so that $G_{\mu\nu}$ {\em must} be interpreted as
gravitational metric. 
There is no need and no room 
for invoking any ``principles''.
Here $\dim \theta = \dim\theta^{\mu\nu} = L^2$, hence
$e^{-\sigma}$ has the dimension $L^{-2n}$ set by the 
noncommutativity scale.
We finally note that 
\be
|G_{\mu\nu}(x)| = |g_{\mu\nu}(x)| , \qquad \mbox{2n=4}
\label{G-g-4D}
\ee
which means that in the 4-dimensional case, 
the Poisson tensor $\theta^{\mu\nu}$ does 
not enter the Riemannian volume at all. This is 
important for stabilizing flat space as we will see,
and is one of several reasons why 4 dimensions are special
in this framework.

\subsubsection{Covariant derivatives and equations of motion.}

To understand the meaning of the 
matrix equations \eq{eom}, we note that the 
double commutator $[X^a, [X^{b},\varphi]] \eta_{ab}$
in the semi-classical limit reduces to 
the covariant Laplacian corresponding
to the effective metric $G_{\mu\nu}$.
This is based on the following facts:

\begin{lemma}
Using the above definitions for 
the metrics
$g_{\mu\nu}, G^{\mu\nu}$ and the (non-degenerate) 
Poisson structure 
$\theta^{\mu\nu}$ on $\cM\subset \R^D$
with Cartesian embedding coordinates  $x^a: \cM \to \R,\,\, a=1,...,D$, 
the following identities hold:
\bea
\{x^a, \{x^{b},\varphi\}\} \eta_{ab} 
&=& e^\sigma \Box_{G} \varphi \label{eom-varphi-0} \\
\nabla^\eta_G (e^{\sigma}\theta^{-1}_{\eta \nu})
&=& G_{\rho\nu}\,\theta^{\rho \mu}
\(e^{-\sigma}\partial_\mu \eta
  + \,\partial_\mu x^a\, \Box_{G} x^b \eta_{ab}\)
\label{theta-covar-id-2}
\eea
Here $\varphi$ is a scalar field
on $\cM\subset \R^D$,   $\nabla_G$
resp. $\Box_{G}$ 
denotes the covariant derivative resp. Laplacian
corresponding to $G_{\mu\nu}$, and
\bea
\eta(x) &:=& \frac 14 e^\sigma \, G^{\mu\nu}(x) g_{\mu\nu}(x) .
\label{eta-def}
\eea
\end{lemma}

\begin{proof}
We first note the  following 
useful identities for the metric  $G^{\mu\nu}$:
\bea
0 &=& 
\partial_\mu (\frac 1{\theta^{n}}\,\theta^{\mu\nu}) 
= \partial_{\mu}(e^{-\sigma}\sqrt{|G|}\,\theta^{\mu\nu}) 
= \sqrt{|G|}\,\nabla_{\mu} (e^{-\sigma} \theta^{\mu\nu}) 
\label{partial-theta-id}\\
\Gamma^\mu &=& - \theta^{n}e^{-\sigma}\,
\partial_\nu (e^\sigma G^{\nu\mu}\,\frac {1}{\theta^{n}})
= - e^{- \sigma}\,\theta^{\nu\nu'}
\partial_\nu (\theta^{\mu\eta} g_{\eta\nu'}(x)) .
 \label{tilde-Gamma}
\eea
The first is a consequence of the Jacobi identity;
for the short proof see Appendix  A.
\eq{tilde-Gamma} follows from \eq{G-def-general} and 
\eq{partial-theta-id}.
Now we compute
\bea
\{X^a, \{X^{b},\varphi\}\} \eta_{ab} 
&=& \theta^{\mu\rho}\partial_\mu x^a
 \partial_\rho(\theta^{\nu\eta}\partial_\nu x^b\partial_\eta\varphi) \,\eta_{ab}\nn\\
&=& \theta^{\mu\rho}\partial_\rho(\partial_\mu x^a
 \theta^{\nu\eta}\partial_\nu x^b\partial_\eta\varphi) \,\eta_{ab}\nn\\
&=& \theta^{\mu\rho}\partial_\rho(\theta^{\nu\eta} g_{\mu\nu}\partial_\eta\varphi) \nn\\
&=& \theta^{\mu\rho}\theta^{\nu\eta} g_{\mu\nu}\partial_\rho\partial_\eta\varphi 
 +  \theta^{\mu\rho}\partial_\rho(\theta^{\nu\eta} g_{\mu\nu})\partial_\eta\varphi \nn\\
&=& e^\sigma ( G^{\rho\eta}\partial_\rho\partial_\eta\varphi-\Gamma^\eta \partial_\eta\varphi )
=  e^\sigma \Box_{G} \varphi ,
\eea
using \eq{partial-theta-id} and \eq{tilde-Gamma}.
Finally,
\eq{theta-covar-id-2}
is shown in  Appendix B \eq{theta-covar-id-2-app},
based on a result in \cite{Steinacker:2008ya}.
\qed 
\end{proof}

Using these identities,  the 
matrix equations of motion 
$[X^b, [X^{a},X^{b'}]] \eta_{bb'} = 0$ in the semi-classical limit become
\bea
\Box_{G} \phi^i &=& 0 ,   \label{eom-phi} \\
\Box_{G} x^\mu &=& 0 
\label{eom-X-harmonic-tree}
\eea
independent of the splitting  
$X^a = (X^\mu,\phi^i)$ into coordinates and scalar fields.
Note that $x^\mu$ is viewed as scalar field on $\cM$ here.
Together with \eq{eom-X-harmonic-tree}, 
\eq{theta-covar-id-2} can be written as
\be
\nabla^\eta_G (e^{\sigma}\theta^{-1}_{\eta \nu})
= G_{\rho\nu}\,\theta^{\rho \mu} e^{-\sigma}\partial_\mu \eta .
\label{theta-covar-id-text}
\ee
These constitute the covariant equations of motion 
of the bare matrix model without matter. 
It is remarkable that \eq{theta-covar-id-text} is completely 
``intrinsic'', i.e. the embedding of $\cM$ plays no role.
It provides the relation between the noncommutativity 
$\theta^{\mu\nu}(x)$ and the metric $G^{\mu\nu}$.
Since \eq{theta-covar-id-text} has essentially the form
of covariant Maxwell equations, it should have a unique solution
for a given ``boundary condition'' such as
\be
\theta_{\mu\nu}(x) \,\to
\, \bar \theta_{\mu\nu} = const 
\qquad \mbox{for}\quad |x|\to\infty 
\label{theta-asympt}
\ee
up to radiational contributions, which will be identified as 
gravitational waves.
We will see that \eq{theta-covar-id-text}
is in fact a consequence of a conservation law resp. Dyson-Schwinger equation,
and is therefore protected from quantum corrections 
\cite{Steinacker:2008ya}.

In general, any $2n$ of the matrices $X^a \sim x^a$ can be considered as 
coordinate functions $x^\mu$ on $\cM$; hence the matrices 
define preferred ``matrix coordinates''.
\eq{eom-X-harmonic-tree} implies that 
these matrix coordinates satisfy the harmonic gauge condition
\be
\Gamma^\mu \,\,\stackrel{\rm e.o.m.}{=} 0  .
\label{gamma-mu}
\ee
From the point of view of GR, \eq{gamma-mu} 
would be interpreted as gauge fixing condition,
disposing of diffeomorphism invariance which does not 
make sense in the matrix model.
Since gauge-dependent objects are always unphysical, this 
has no implications on the physical content of the model.
The equations of motion in these matrix coordinates
are given in  \cite{Steinacker:2008ri}.

Finally a comment on Lorentz invariance is in order. 
Even though $\theta^{\mu\nu}$ explicitly 
breaks Lorentz invariance (just like any other background
field), the physical fields of the model
do not couple directly to  $\theta^{\mu\nu}$, and 
the effective actions such as 
\eq{covariant-action-scalar}, \eq{action-expanded-2}
are (locally) Lorentz-invariant 
with respect to 
$G_{\mu\nu}$. There is no field in the 
model which is charged under the trace- $U(1)$ gauge field 
whose flux corresponds to 
$\theta^{-1}_{\mu\nu}$. Therefore 
Lorentz-breaking effects can arise only through higher-order
corrections\footnote{In fact the next-to-leading
order corrections are expected to 
be suppressed by $O(\theta^2)$, because 
one can choose a star product \cite{Kontsevich:1997vb} such that 
$[f,g] = i\{f,g\} + O(\theta^3)$.} 
in $\theta^{\mu\nu}$ or through quantum effects,
both of 
which are expected to be suppressed by powers of 
$\L_{NC}^{-2}$, or probably $\L_{NC}^{-4}$.

\subsection{Effective action, degrees of
freedom and variational principle}
\label{sec:dof}

We now show how to obtain the above geometrical
equations of motion directly from the semi-classical limit of the
matrix model action \eq{YM-action-1}, which 
using \eq{rho-sigma-det}, \eq{eta-def}
can be written as 
\be
S = - \frac{\Lambda_0^4}{4 g^2}\, \Tr [X^a,X^b][X^{a'},X^{b'}] 
\eta_{aa'} \eta_{bb'} 
\, \sim \,  \frac{\Lambda_0^4}{(2\pi)^n g^2}\,\int d^{2n} x\, \sqrt{|G|}\,e^{-\sigma}\eta .
\label{S-semiclassical-general}
\ee
Before applying the variational principle, we should
first clarify the independent degrees of freedom.

\paragraph{Geometrical degrees of freedom.}

The semi-classical action \eq{S-semiclassical-general} 
depends on the Poisson tensor $\theta^{\mu\nu}(x)$ and the embedding
metric $g_{\mu\nu} = \eta_{\mu\nu} + \del_\mu\phi^i\del_\nu\phi^j\d_{ij}$,
which in turn depends on the embedding functions $\phi^i(x^\mu)$. 
On the other hand, the fundamental degrees of freedom are the 
$D$ matrices resp. functions $X^a$, which can be 
varied independently.
Semi-classically, these variations $\d X^a$ can be decomposed into 
tangential and normal fluctuations w.r.t. 
the background brane $\cM$. This is most transparent in 
normal embedding coordinates \eq{normal-embedd-coords}, where
$\d\phi^i \in T\cM^\perp$ resp. $\d X^\mu \in T\cM$
correspond to transversal resp. tangential variations
of the brane.
Clearly the variations of the scalar fields $\phi^i(x)$ 
change the embedding $\cM \subset \R^D$ and 
the embedding metric $g_{\mu\nu}$, while tangential
variations 
\bea
X^\mu &\to& (X^\mu)' = X^\mu + \cA^\mu 
= X^\mu - \theta^{\mu\nu}(x) A_\nu, \nn\\
x^\mu &\to& (x^\mu)' = x^\mu + \cA^\mu 
\label{A-diffeo}
\eea
lead to a variation of the Poisson tensor on $\cM$  according to 
\bea
\theta^{\mu\nu} \sim -i [X^\mu,X^\nu] 
&\to& -i [(X^\mu)',(X^\nu)']  \sim \theta^{\mu\nu} 
+ \theta^{\mu\rho}\partial_\rho \cA^{\nu}
- \theta^{\nu\rho}\partial_\rho \cA^{\mu} -i [\cA^{\mu},\cA^{\nu}] 
\nn\\
&=&  (1+\cA\cdot \partial)\theta^{\mu\nu} 
 - \theta^{\mu\mu'} \theta^{\nu\nu'}F_{\mu'\nu'}\nn\\
&=& (\theta^{\mu\nu})'
\label{theta-trafo}
\eea 
using the Jacobi identity, where
$F_{\mu'\nu'} = \partial_{\mu'} A_{\nu'} 
- \partial_{\nu'} A_{\mu'}$ and
$\cA\cdot \partial \equiv \cA^\rho\partial_\rho$.
For the symplectic 2-form, this amounts to 
\be
(\theta^{-1}_{\mu\nu})' = \theta^{-1}_{\mu\nu}-
\theta^{-1}_{\mu\mu'} \theta^{-1}_{\nu\nu'} \d\theta^{\mu\nu}
= (1+\cA\cdot \partial) \theta^{-1}_{\mu\nu} + F_{\mu\nu} .
\label{omega-variation}
\ee
Note that these horizontal variation 
determine a vector field $\cA = \cA^\mu \partial_\mu$ and thus 
an (infinitesimal) diffeomorphism $\cM \to \cM$.
In particular, the $U(\infty)$ gauge symmetry \eq{gauge} 
resp. its infinitesimal version
$X^a \to (X^a)' = X^a - i[\lambda,X^a]$
and $\phi^i \to \phi^i - i[\lambda,\phi^i]$
corresponds to
\be
\d X^\mu = \cA^\mu 
\sim -\theta^{\mu\nu}\partial_\nu \lambda(x) ,
\qquad \d \phi^i 
\sim -\theta^{\mu\nu}\partial_\mu \phi^i\partial_\nu \lambda(x).
\ee
This defines  an exact symmetry of the matrix model corresponding
to (infinitesimal) {\em symplectomorphisms} ${\rm Symp_\theta}$
\be
\{\l,.\} = (\partial_\nu\lambda)\theta^{\nu\mu} \partial_\mu .
\label{symplect}
\ee
By definition,  symplectomorphisms 
leave $\theta^{\mu\nu}(x)$ invariant
$\cL_{\cA_\lambda} \omega = 0$, and form a
subgroup\footnote{It is in general much smaller that the 
volume-preserving diffeos except in 2 dimensions, hence a
comparison with unimodular gravity is quite inappropriate.}
of the group of volume-preserving diffeomorphisms
due to Liouvilles theorem.
From the point of view of $U(1)$ gauge theory on $\R^4_\theta$, 
they correspond to gauge transformations 
$A_\mu \to A_\mu + \partial_\mu \lambda(x)$, thus providing 
a connection between gauge theory and (emergent) gravity.

A more geometric understanding is obtained as follows: 
since locally the $X^\mu$ generate the full matrix algebra, a general variation of 
the matrices ${X^a}' = X^a + \cA^a \sim x^a(x^\mu) + \cA^a(x^\mu)$ corresponds
semi-classically to a modified embedding 
\be
{X^a}': \,\, \cM'\, \stackrel{\Psi}{\rightarrow} \,\cM \,\hookrightarrow\, \R^D
\label{variation-geom}
\ee 
in terms of some map $\Psi$, i.e. $(x^a)' = x^a \circ \Psi$. 
In general $\Psi$ will be neither an isometry nor a symplectomorphism, and the 
new configuration ${X^a}'$ is completely inequivalent to $X^a$. If $\Psi$
happens to be a diffeomorphism $\cM\to \cM$ (in particular it preserves the embedding), 
then the induced metric $g_{\mu\nu}$ is equivalent, but the Poisson structure 
and therefore $G^{\mu\nu}$ is not.
This is the case of tangential transformations \eq{A-diffeo}. 
In particular, gauge transformations \eq{gauge} define such a
symplectomorphism
$\Psi$ on $\cM$, hence $\{x^a\circ\Psi,x^b\circ\Psi\} = \{x^a,x^b\}\circ\Psi$,
and the effective geometry is the same.
The matrix model action and 
its semi-classical limit are invariant under this symmetry,
because the symplectic volume resp. the trace is preserved.

Hence in the matrix model, transformations should always be considered as 
``active'' as in \eq{variation-geom}, 
rather than as diffeos in some given coordinate system.
All  $x^a$ behave as scalar fields on $\cM$ and are on equal footing.
This is why gauge transformations \eq{gauge}
(and more generally the transformations \eq{A-diffeo}) 
do not appear to act as Lie derivatives;
for example, \eq{omega-variation} is {\em not} the Lie derivative 
of a 2-form along the vector field $\cA$ which would be  
$\cL_\cA \omega =  F_{\mu\nu} dx^\mu\wedge dx^\mu$.

The bottom line is that the 4 tangential degrees of freedom $\cA^\mu$
are transmuted into the 3 (resp. 2 ``on-shell``)  degrees of freedom of 
$\omega = \frac 12 \theta^{-1}_{\mu\nu} dx^\mu\wedge dx^\nu$ 
resp. the $U(1)$ field strength 
$F_{\mu\nu}$, plus {\em one}  gauge degree of freedom
corresponding to symplectomorphisms. 
This means that one can consider 
the embedding $\phi^i$ and $\theta^{\mu\nu}$ as independent 
geometrical
degrees of freedom, and use a semi-classical variational 
principle for the actions of the type
\be
S = \int d^{2n} x\,\rho \cL(\phi^i(x),\theta^{\mu\nu}(x))
\ee
as illustrated below.
Even though this gives the correct equations of motion, 
one should keep in mind that the fundamental
degrees of freedom are the matrices $X^a$. Their variation
does not simply separate into $\d\theta^{\mu\nu}$ 
and $\d \phi^i$ but involves
also a change of coordinates.

\paragraph{Semi-classical  derivation of the equations of motion.}

Once the classical 
degrees of freedom are understood, one can work with
the effective action \eq{S-semiclassical-general} for the matrix model
in terms of a sub-manifold $\cM \subset \R^D$,
and derive covariant equations of motion for the embedding and the Poisson 
structure. As we just explained,
the most general variations of the latter
can be parametrized in terms of a $U(1)$ gauge field as
\be
\d \theta^{-1}_{\mu\nu} = 
\nabla_\mu \d A_\nu - \nabla_\nu \d A_\mu ,
\ee
and the variation of the embedding can be captured by
$\d\phi^i$.
Thus the variation of the effective action 
\eq{S-semiclassical-general} is
\bea
\d S &=& \int  d^{2n}x\, \sqrt{|\theta^{-1}_{\mu\nu}|}\Big(\d \eta(x)\,  
+ \frac 12 \theta^{\mu\nu}\d \theta^{-1}_{\nu\mu} \Big)\nn\\ 
&=& \frac 12 \int  d^{2n}x\, \sqrt{|\theta^{-1}_{\mu\nu}|}
\Big(g_{\mu\nu} \theta^{\mu\mu'}\d\theta^{\nu\nu'}g_{\mu'\nu'} \,  
+ g_{\mu\nu} \theta^{\mu\mu'}\theta^{\nu\nu'}\d g_{\mu'\nu'} \, 
+ \,\eta(x)\,\theta^{\mu\nu}\d \theta^{-1}_{\nu\mu} \Big)\nn\\
&=& \frac 12\int  d^{2n}x\,\sqrt{|\theta^{-1}_{\mu\nu}|}
\Big(e^{2\sigma} G^{\eta\mu}\theta^{-1}_{\mu\nu}G^{\nu\rho}
\d\theta^{-1}_{\rho\eta}\,
+ e^{\sigma} G^{\mu\nu} \d g_{\mu\nu}
+ \,\eta(x)\,\theta^{\mu\nu}\d \theta^{-1}_{\nu\mu} \Big)\nn\\
&=& \int  d^{2n}x\,\sqrt{|G|} 
\Big(G^{\eta\mu} G^{\nu\rho} 
e^{\sigma}\theta^{-1}_{\mu\nu}\nabla_\rho \d A_\eta\,
- \,e^{-\sigma}\eta\,\theta^{\rho\eta} 
\nabla_\rho \d A_\eta\,  
+ G^{\mu\nu} \partial_\mu \phi^i  \partial_\nu \d\phi^i\, \d_{ij} \Big) 
\nn
\eea
using \eq{rho-sigma-det}.
Using partial integration
$\int d^{2n}x\, \sqrt{|G|}\, \nabla_\mu V^\mu =0$
and $\nabla G =0$ 
we obtain
\bea
\d S &=& -\int  d^{2n}x \, \sqrt{|G|} \,\d A_\eta 
\Big(G^{\eta\mu}G^{\nu\rho}
\nabla_\rho(e^{\sigma} \theta^{-1}_{\mu\nu})\,
- \,\nabla_\rho(e^{-\sigma}\eta\,\theta^{\rho\eta}) \Big)
 + \d\phi^i\,\d_{ij} \partial_\nu 
\(\sqrt{|G|}\, G^{\mu\nu}\partial_\mu \phi^i \) \nn\\
&=& -\int  d^{2n}x\, \sqrt{|G|} \,\(\d A_\eta 
\Big(G^{\eta\mu} G^{\nu\rho}
\nabla_\rho(e^{\sigma} \theta^{-1}_{\mu\nu})\,
-\,|G|^{-1/2} \partial_\rho 
(|G|^{1/2}e^{-\sigma}\eta \theta^{\rho\eta})\Big) 
 + \d\phi^i\,\d_{ij} \Box_{G} \phi^i \) \nn\\
&=& -\int  d^{2n}x\,\sqrt{|G|} \,\(\d A_\eta 
\Big(G^{\eta\mu}G^{\nu\rho}
 \nabla_\rho(e^{\sigma} \theta^{-1}_{\mu\nu})\,
-e^{-\sigma}\theta^{\rho\eta}\,\partial_\rho \eta \Big)
 + \d\phi^i\,\d_{ij} \Box_{G} \phi^i\)\, \nn
\label{variation-bare-geometry}
\eea
using \eq{partial-theta-id} and \eq{rho-sigma-det} in the last steps.
This gives precisely the equations of motion 
\eq{theta-covar-id-text} and \eq{eom-phi}.
Comparing with \eq{theta-covar-id-2}, it may seem puzzling 
that $\Box_{G} x^\mu =0$ 
 must be assumed as well
in order to get \eq{theta-covar-id-text}.
This can be understood as follows (assuming $g=G$): Since the 
 $x^a$ are Cartesian coordinates on $\R^D$, 
$\Box_{G} x^\mu =0$  is actually a consequence of  
$\Box_{G} \phi^i =0$, expressing the fact that $\cM \subset \R^D$
is a minimal surface. Thus from the GR point of view, $x^\mu$ is gauge fixed. The physical degrees
of freedom in $X^\mu$ are transmuted into $\theta^{\mu\nu}$,
which satisfies the independent equation of motion \eq{partial-theta-id}.

Finally the propagation of the $U(1)$ degrees of freedom contained in 
$\theta^{\mu\nu}$ resp. $\d A_\mu$ is governed by a Maxwell-like action, 
which for flat embeddings is precisely the $U(1)$ gauge theory discussed
above, and for nontrivial backgrounds takes the same form\footnote{this
was elaborated by A. Schenkel, unpublished} as 
for the nonabelian components 
discussed in section \ref{sec:nonabelian-gauge}.

\paragraph{Remark on extra dimensions.}

The discussion so far applies to noncommutative spaces
$\cM^{d}$ with arbitrary even dimension $d=2n$. 
Even though the primary interest is on 4-dimensional 
space-time, this is important because any realistic 
scenario is likely to involve extra dimensions, 
such as $\cM^{d} = M^4 \times K^{2m} \subset \R^{10}$ 
where $K^{2m}$ is some small compact space. 
We will see that the nonabelian gauge fields
obtained on $\cM^{d}$ are a priori always $\msu(N)$-valued,
which must be broken spontaneously 
to obtain some more realistic gauge group.
A natural way to achieve this is through (fuzzy)
extra dimensions, cf.  
\cite{Aschieri:2006uw,Grosse:2010zq,Chatzistavrakidis:2010xi}.
Such extra dimensions may also play an important role in the 
quantization and will affect the scaling of various parameters such
as the effective gravity scale. These are topics for further research.

\paragraph{Relation with string theory.}

The IKKT matrix model was proposed originally as a non-perturbative
definition of type IIB string theory on $\R^{10}$. Our results are indeed
very reminiscent of D-branes in a $B$-field background 
\cite{Seiberg:1999vs}:
$\cM_\theta$ could be interpreted as a brane in $\R^{10}$
with open string metric $\sim G_{\mu\nu}$, while
$g_{\mu\nu}$ corresponds to the closed string metric, and
$\theta^{-1}_{\mu\nu}$ is essentially the $B$ field (more precisely
$B+F$, absorbing the $U(1)$ gauge field). There are also other 
solutions of the matrix model consistent with this interpretation, 
such as superpositions of branes. 
Graviton scattering has been studied from this point of view 
in  matrix models, and a relation with supergravity
has been conjectured; for an
incomplete list of references see 
\cite{Banks:1996vh,Banks:1996nn,Ishibashi:1996xs,Chepelev:1997ug,Aoki:1998vn,Aoki:1999vr,Nair:1998bp,Nishimura:2001sx,Taylor:2001vb,Kitazawa:2006pj}
and references therein.
However, most of this work was focused on the interaction of different block-matrix configurations, 
thereby probing the bulk metric and the relation with 10D supergravity. NC branes have been considered only
for geometries with a high degree of symmetry. 
The essential new point here is to consider 
{\em curved} NC branes in the matrix model and to study
the effective gravity on such a 4-dimensional brane solution, which does not require any 10D physical bulk. 
This is indeed possible, and it opens the possibility to address  
4-dimensional gravity through this simple type of matrix model.
Thus the strength of string theory  
(notably the good behavior under quantization) 
seems preserved while the main problems
(lack of predictivity) are avoided.
In particular, the matrix model should be viewed as
background-independent, since 
physical space-time emerges dynamically.

\paragraph{Related aspects and results.}

The topic of membranes in matrix models
has a long history, cf. \cite{deWit:1988ig,Nicolai:1998ic,Hoppe:2002km}. 
There has been extensive recent work
on the noncommutative geometry and physics of such 
``special'' NC branes, starting with the 
fuzzy sphere \cite{Madore:1991bw}, which generalizes to fuzzy $\C P^n$
\cite{Grosse:2004wm}
and to arbitrary quantized coadjoint orbits,
$q$-deformed versions \cite{Grosse:2000gd}, etc..
Other examples include the fuzzy torus \cite{Ambjorn:2000cs}, and 
various examples with less or no symmetry, cf. 
\cite{Arnlind:2006ux,Cornalba:1998zy} 
and references therein. 
These and similar spaces may play a physical role
as compactified extra dimension.
Matrix models with additional (quadratic or cubic) terms are known to 
admit such homogeneous spaces as solutions, cf. 
\cite{Nair:1998bp,Myers:1999ps,Iso:2001mg,Azuma:2004qe,Alekseev:2000fd,
Grosse:2004wm,DelgadilloBlando:2008vi,Yang:2009pm}.

The present framework can be seen as a realization of the ideas in 
\cite{Rivelles:2002ez}, where NC $U(1)$ gauge fields
were interpreted as gravitons;
this will be explained in section \ref{sec:linearized-U(1)}.
A relation between noncommutative gauge theory and 
geometry resp. emergent gravity
has also been advocated in \cite{Yang:2004vd,Yang:2006mn,Yang:2008fb}, 
relating in particular self-dual NC
Maxwell theory with self-dual gravity. These general ideas are clearly similar  
in spirit and related to the results presented here, although the precise relation
is not always clear to the author.
Here we restrict ourselves to the case of matrix models, which shows
the need to consider
nontrivially embedded branes as well as quantum effects.
It is well-known that such matrix models are very rich and admit
geometrical phenomena (change of dimension, orientifolds, etc.), 
see e.g. \cite{Sochichiu:2005ex,Itoyama:1998et}.
For other related work see e.g. 
\cite{Banerjee:2004rs,Muthukumar:2004wj}, and \cite{Chaichian:2006ht}
for a different approach to NC surfaces.

\subsection{$SU(n)$ gauge fields coupled to gravity} 
\label{sec:nonabelian-gauge}

To avoid notational conflicts, 
we denote the basic matrices 
with $Y^a$ in this section, governed by 
the same matrix model as above
\be
S_{YM} = - \frac{\L_0^4}{4 g^2} Tr [Y^a,Y^b] [Y^{a'},Y^{b'}] 
\eta_{aa'}\eta_{bb'} \, .
\label{YM-MM-nonabel}
\ee
However we now consider a matrix background corresponding to $n$ coinciding branes
\be
Y^a = \left(\begin{array}{l}Y^\mu \\ Y^i \end{array}\right) = 
\left\{\begin{array}{ll} X^\mu\otimes \one_n,  & \quad a=\mu = 1,2,..., 2n,   \\ 
\phi^i \otimes \one_n, & 
\quad a = 2n+i, \,\, i = 1, ..., D-2n . \end{array} \right.
\ee
We want to understand general fluctuations around this background.
Since the $U(1)$ components describe the geometry, 
we expect to find $su(n)$-valued gauge fields as well as
 scalar fields in the adjoint.
It turns out that the following gives a useful parametrization 
of these general fluctuations:
\be
\left(\begin{array}{l}Y^\mu \\ Y^i \end{array}\right) = 
\left(\begin{array}{l} X^\mu\otimes \one_n + \cA^\mu   \\ 
(1+ \cA^\nu\partial_\nu)\phi^i\otimes \one_n + \Phi^i
\end{array} \right)
\sim 
(1+ \cA^\nu\partial_\nu)\left(\begin{array}{l}Y^\mu \\ Y^i \end{array}\right) +
\left(\begin{array}{l}0  \\ 
 \Phi^i\end{array} \right)
\label{nonabelian-SW}
\ee
where\footnote{We drop sub-leading terms, such as
$\frac 12 \{\cA^\rho, \partial_\rho \Phi^i\}$ for the nonabelian scalars.}
\bea
\cA^\mu &=&  \cA^\mu_\a \otimes \lambda^\a
\,\, = \,\, - \theta^{\mu\nu} A_{\nu,\a} \otimes \lambda^\a, \nn\\
\Phi^i  &=&  \Phi^i_\a\otimes \l^\a 
\eea
parametrize the $su(n)$-valued gauge fields resp. scalar fields, and
$\lambda^\a$ denotes the generators of $su(n)$.
This amounts to the 
leading term in a Seiberg-Witten (SW) map 
\cite{Seiberg:1999vs}, which relates
noncommutative and commutative gauge theories with the appropriate 
gauge transformations. 
This SW parametrization can be characterized by requiring that the 
noncommutative gauge transformation 
$\d_{nc} Y^a = i [Y^a,\Lambda]$ 
induces for the $su(n)$ components
$A_\mu = A_{\mu,\a} \la^\a$ and $\Phi^i$
the ordinary gauge transformations 
\bea
\d_{cl} A_\mu &=& i[A_\mu,\Lambda]_{su(n)} \, +\, \partial_\mu\Lambda(x) 
\quad + O(\theta)\nn\\
\d_{cl} \Phi^i &=& 
i [\Phi^i,\Lambda]_{su(n)} \quad + O(\theta^2) ,
\eea 
cf. \cite{Jurco:2000fb}.
Here the subscript $[\Phi^i,\Lambda]_{su(n)}$
indicates that only the commutator of the explicit $su(n)$ generators
is to be taken, but not the $O(\theta)$ contributions from the 
Poisson bracket. This strongly suggests that the matrix model action
expressed in terms of these $A_\mu$ should reduce to a conventional
gauge theory in the semi-classical limit;
 this was verified in \cite{Steinacker:2007dq}.

In our context, the SW map \eq{nonabelian-SW} can be understood geometrically
in a very simple way. Recall that the $U(1)$ sector  
(i.e. the components
proportional to $\one_n$) describes the
geometrical degrees of freedom:  
in the geometrical limit,
$X^a = (X^\mu,\phi^i)$ become functions
\be
x^a = (x^\mu,\phi^i(x))
\ee
on $\cM$ which describe the embedding of the $2n$-dimensional brane
$\cM \subset \R^D$. Then
the one-form $A_\mu dx^\mu$ together with the 
Poisson tensor determines a tangential vector field
\be
A_\mu e^\mu = A_\mu\theta^{\mu\nu}\partial_\nu = \cA^\nu \partial_\nu
\quad \in T_p\cM \otimes \msu(n),
\ee 
whose push-forward in the ambient space $\R^D$ 
\be
\cA^\nu\partial_\nu x^a
\cong \cA^\nu(\d_\nu^\mu,\partial_\nu \phi^i)
\ee
coincides with the fluctuations $\d X^a = Y^a - X^a$ 
of the dynamical matrices in \eq{nonabelian-SW} 
(for vanishing $\Phi^i$).
This provides the link between gauge fields 
and ``covariant coordinates'' $X^a$.
The nonabelian scalar fields $\Phi^i$
can be thought of as coordinate functions
 of $\cM$ embedded in further 
extra dimensions corresponding to $su(n)$. They
behave as scalar fields, but might contribute to the
background geometry if they acquire a non-trivial VEV,
completely analogous to the $U(1)$ components $\phi^i$.

After this preparation, 
we can write down the effective action for $su(n)$-valued
gauge fields $A_{\mu}$ 
on general $\cM_\theta^{2n}\subset \R^D$
in the matrix model \eq{YM-MM-nonabel}
in the semi-classical limit:
\bea
S_{YM}[\cA] &\sim&
\frac{\Lambda_0^4}{4 g^2}\int d^{2n} x\, \sqrt{|G_{\mu\nu}|}
e^{\sigma} \, G^{\mu\mu'} G^{\nu\nu'} 
\tr(F_{\mu\nu}\,F_{\mu'\nu'})\,\, 
-\,\, S_{NC} 
\label{action-expanded-2}
\eea
where $F_{\mu\nu} = \partial_\mu A_\nu - \partial_\nu A_\mu + i
[A_\mu,A_\nu]$
is the $su(n)$-valued field strength, 
and
\bea
S_{NC} &=& \frac{\Lambda_0^4}{4 g^2} \int d^{2n}x\,\sqrt{|\theta^{-1}_{\mu\nu}|}
 \Big( F_{\mu'\nu'} F_{\mu\nu} \hat\theta^{\mu'\nu'} \theta^{\mu\nu} 
+ 2 F_{\mu'\mu} F_{\nu'\nu}\hat\theta^{\mu'\nu'}\theta^{\nu\mu}\nn\\  
&& \qquad - \frac 12 \eta\theta^{\mu\nu}\theta^{\mu'\nu'}
\big(F_{\mu\nu}F_{\mu' \nu'} + 2 F_{\mu\nu'} F_{\nu\mu'} \big)  \Big) \nn\\
&\stackrel{n=2}{=}& - \frac{\Lambda_0^4}{2 g^2} \int  \eta(x)\,\tr F\wedge F \; .
\label{S-NC-def}
\eea
The last line holds for 4-dimensional branes, and can be seen using
\be
\frac 12 (F\wedge F)_{\mu\nu\rho\s} \hat\theta^{\mu\nu}\theta^{\rho\s}
= (F_{\mu\nu}\hat\theta^{\mu\nu}) (F_{\rho\s}\theta^{\rho\s})
 +  2 F_{\mu\s} F_{\nu\rho} \hat\theta^{\mu\nu}\theta^{\rho\s} 
\label{FF-theta}
\ee
and $\hat\theta\wedge\theta = \eta e^{-\sigma} \theta\wedge\theta$, see
\cite{Steinacker:2007dq}.
Note that $g_{\mu\nu}$ enters the ``would-be topological term'' $S_{NC}$
through $\eta$ resp. the antisymmetric matrix
\be
\hat \theta^{\nu\eta} 
= G^{\nu\rho} g_{\rho\mu}(y)\theta^{\mu\eta} 
= - e^\sigma\,G^{\nu\rho} \theta^{-1}_{\rho\mu} G^{\mu\eta} 
= -\hat \theta^{\eta\nu} .
\label{theta-hat}
\ee
 This result is non-trivial;
for the 4-dimensional matrix model it was first obtained
through a direct but rather non-transparent
computation of the action \cite{Steinacker:2007dq}, 
which requires the 2nd order Seiberg Witten map.
For the general case a proof  based on 
the conservation law \eq{e-m-cons} was given
\cite{Steinacker:2008ya}, which is not only much
simpler but also establishes the corresponding 
Yang-Mills equations of motion at the quantum level.

It follows that the (bare) $\msu(n)$ Yang-Mills coupling constant
is given by 
\be
g^2_{YM} = \frac{e^{-\sigma}}{\Lambda_0^4}\,   g^2 .
\ee
The factor $e^\sigma$ is reminiscent of a dilaton, however it is not an
independent field here but determined by the geometry. 
This implies that $e^{-\sigma}$ must be nearly constant
in any realistic solution, which in turn implies that
$e^{-\sigma} \sim \L_{NC}^2$ must be nearly constant.
In the case of a (nearly-) 
self-dual symplectic form as discussed
in section \ref{sec:self-dual}, we have $\eta = e^\sigma$
and 
\bea
S_{YM}[\cA]  &=& 
\frac{\Lambda_0^4}{4 g^2} \int 
 \,e^{\sigma} \tr\(d^{4} x\, |G_{\mu\nu}|^{1/2}\,G^{\mu\mu'} G^{\nu\nu'}
F_{\mu\nu}\,F_{\mu'\nu'}\,\, + \, 2  F\wedge F \)  \; .
\label{action-expanded-2-SD}
\eea
which is essentially a selfdual combination.
Recall also that the $A_{\mu}$ are $\msu(n)$
gauge fields, hence any realistic gauge theory must descend 
via some symmetry breaking mechanism
such as in \cite{Chatzistavrakidis:2010xi} before phenomenological 
conclusions can be drawn.

\paragraph{Nonabelian scalars.}

It follows easily along the lines of section \ref{sec:emergent-geom} that
the geometrical action for the $su(n)$ -valued scalars 
$\Phi^i_\a$ in the matrix model at leading order is
\bea
S_{YM}[\Phi] &\sim& \frac{\L_0^4}{2 (2\pi)^n g^2}\int d^{2n} x\, \sqrt{|G|}\,
G^{\mu\nu} \tr (D_\mu \Phi^i  D_\nu \Phi^i   
+ \frac 12 e^{-\sigma}\, 
[\Phi^i ,\Phi^j ][\Phi^{i'} ,\Phi^{j'} ] \d_{ii'} \d_{jj'}) , \nn\\
\label{S-phi-effective-nonabelian}
\eea
where $D_\mu \Phi^i = \partial_\mu \Phi^i + i [A_\mu,\Phi^i]$. 
It is worth pointing out that in the case $D=10$ and upon
adding suitable fermions as in the IKKT model \cite{Ishibashi:1996xs}, 
this will lead to the analog of $N=4$ SYM theory on 
$\R^4$, or on a more general geometry. The SUSY transformations in general
act non-trivially on the geometry, 
as indicated below \cite{Klammer:2008df}.

One might question why only the $U(1)$ components should be 
considered as geometrical degrees of freedom, in contrast 
to the $SU(n)$ components. 
The special role of the trace-$U(1)$ components lies 
in the fact that they couple universally via commutators
to all other fields. This universal coupling is responsible
for gravity, as shown explicitly. In contrast, the $SU(n)$
components of nonabelian gauge fields  
couple primarily through their nonabelian nature.

\subsection{Fermions}
\label{sec:fermions}

The most obvious (perhaps the only reasonable) action
for a spinor which can be written down in the matrix 
model framework\footnote{In particular, 
fermions should also be in the adjoint, otherwise they
cannot acquire a kinetic term. This is not incompatible
with particle physics, see e.g. 
\cite{Chatzistavrakidis:2010xi,Grosse:2010zq}.}
is 
\bea
S[\Psi] &=& \Tr\, \obar\Psi  \Gamma_a [X^a,\Psi]
\sim \,\, \frac 1{(2\pi)^{n}} \int d^{2n} x\, \rho(x)\, \obar\Psi i 
\gamma_\mu
 \theta^{\mu\nu}(x) D_\nu\Psi   \nn\\
&=& \,\, \frac 1{(2\pi)^{n}} \int d^{2n} x\, \rho(x)e^{\sigma/2}\, 
\obar\Psi\, i \tilde\gamma^\mu D_\mu\Psi  ,
\label{fermionic-action-geom}
\eea
dropping the nonabelian scalars $\Phi^i$ for simplicity (see below).
This should be added to the bosonic action \eq{YM-action-1}.
The  Euclidean version involves the obvious
replacement $\bar\Psi \to \Psi^\dagger$.
Here $\Gamma_a$ defines the D-dimensional Clifford
algebra. We defined the ``local Clifford generators''
\be
\gamma_\mu(x) = \del_\mu x^a \Gamma_a,  \qquad 
\tilde \gamma^\mu(x) = e^{-\sigma/2}\,\gamma_\nu\, \theta^{\nu\mu},
\label{tilde-gamma}
\ee
which
satisfy the Clifford algebra associated with the metrics
$g_{\mu\nu}(x)$ and $G^{\mu\nu}(x)$ on $\cM$,
\bea
\{\gamma_\mu, \gamma_\nu\} &=& 2 g_{\mu\nu}(x) \, , \nn\\
\{\tilde \gamma^\mu,\tilde\gamma^\nu\} &=&  2 \, G^{\mu\nu}(x)\, .
\eea
Thus the (matrix) Dirac operator can be written as
\be
\slashed{D}\Psi = \Gamma_a \left[X^a, \Psi\right] 
\sim  i\tilde\gamma^\mu D_\mu \Psi.
\label{eq:Dirac op}
\ee
As pointed out in~\cite{Klammer:2008df}, $\slashed{D}\Psi$ does not quite 
match with the standard covariant Dirac operator for spinors
$\slashed{D}_{\mathrm{comm}}\Psi= i \gamma^a e_a^\mu \left(\partial^\mu + \Sigma_{bc}\omega^{bc}_\mu\right)\Psi$,
where
$\omega^{ab}_\mu = \frac{i}{2}e^{a\nu}\nabla_\mu e^b_\nu$
is the usual spin connection, and
$\Sigma_{ab}=\frac{i}{4}[\gamma_a,\gamma_b]$. 
While the explicit derivative term in \eq{eq:Dirac op}
is essentially the same, 
the spin connection is trivial\footnote{Incidentally, the present 
Dirac operator on submanifolds is the analog of the operator used 
in Witten's proof of the positive energy theorem 
\cite{Witten:1981mf}.}  in the matrix coordinates $x^\mu$.
This is consistent because the spinors are those of the ambient space $\R^D$. 
To understand this in more physical terms,
recall that the spin connection determines how the spinors are rotated 
under parallel transport along a trajectory.
However, $\omega^{ab}_\mu$ can always be eliminated 
(via a suitable gauge choice)
along an open trajectory, so that the conventional
kinetic term essentially coincides \eq{eq:Dirac op}
in the point-particle limit. Hence
the trajectory of a classical fermion with action
\eq{fermionic-action-geom} 
will follow properly the geodesics 
of the metric\footnote{for massless particles, the geodesics 
are independent of possible conformal factors.
Masses should be generated spontaneously, which is not studied 
here.} $G_{\mu\nu}$, 
albeit with a non-standard gravitational ``spin-dragging''.
This means that holonomies here
will be different than in General Relativity, and the gravitational 
spin rotation for a free-falling fermions might provide
a characteristic signature for the framework of matrix models.
For special geometries, $\slashed{D}$ coincides with
the usual Dirac operator on curved manifolds, e.g. in the case
of the fuzzy sphere \cite{Grosse:1994ed}.

We will see below that upon integrating out the fermions, 
an induced gravitational action is obtained which
induces the expected Einstein-Hilbert term
$\int d^4 x\,  \L^2\, R[G]\,$, albeit with an 
unusual numerical coefficient, as well as additional terms. 
All this shows that \eq{fermionic-action-geom} defines a
reasonable action for fermions in the background defined by 
$G_{\mu\nu}$.

\paragraph{Supersymmetry.}

The presence of fermions in the matrix model is important
not only to describe physical fermions but also because they
allow supersymmetry. The IKKT model \cite{Ishibashi:1996xs}
\be
S_{IKKT} = - \frac{\Lambda_{0}^4}{g^2}\, \Tr \( \frac 14 [X^a,X^b] [X^{a'},X^{b'}] 
\eta_{aa'}\eta_{bb'} \,\, + \,\, \frac 12\, \obar\Psi  \Gamma_a
[X^a,\Psi]\) 
\ee
with $D=10$ and 
$\Psi$ a Majorana-Weyl spinor is singled out by the
existence of the following maximal supersymmetry  
\bea
\delta^{(1)} \Psi &=& \frac i2 [X^a,X^b] \,\Gamma_{ab} \epsilon, 
\qquad \,
\delta^{(1)} X^a = i \bar \epsilon \Gamma^a \Psi, \nn\\
\delta^{(2)} \Psi &=& \xi, \qquad \qquad  \qquad\qquad 
\delta^{(2)} X^a = 0 ,
\label{susy-trafo-1}
\eea
where $\Gamma_{ab} =\frac 12 [\Gamma_a,\Gamma_b]$,
and $\epsilon,\xi$ are Grassmann-valued spinors.
This model is obtained by dimensional reduction of 
$\cN=1$ super-Yang-Mills theory on $\R^{9,1}$ to a point.

To recover spacetime supersymmetry, 
we split the matrices again into 4 + 6 dimensions $X^a = (X^\mu,\phi^i)$.
Then the 4-dimensional Moyal-Weyl quantum plane $\R^4_\theta$
is a (BPS) solution of the
generalized matrix equations of motion, embedded as
\bea
X^\mu &=& \bar X^\mu, \qquad \mu= 0,...,3, \nn\\
\phi^i &=& 0\, .
\eea
All previous geometrical
considerations can be generalized,
except that the matrix model now contains scalar fields $\phi^i(x)$.
For simplicity we only discuss the case of $\R^4_\theta$ 
with flat embedding $\del_\mu\phi^i = 0$.
If we set 
$\xi = \frac 12 \bar\theta^{\mu\nu} \Gamma_{\mu\nu} \epsilon$
following \cite{Ishibashi:1996xs} 
and recall $X^\mu = \bar X^\mu - \bar\theta^{\mu\nu} A_\nu$
and \eq{tilde-gamma},
then the combined transformation 
$\delta = \delta^{(1)} + \delta^{(2)}$
takes the form
\bea
\delta \Psi
&=& -\Big( i e^\sigma F_{\mu\nu} \tilde\Sigma^{\mu\nu}
 + e^{\sigma/2} D_\mu\Phi^i \tilde\gamma^\mu\gamma_5\Delta_i
 -\frac i2[\Phi^i,\Phi^j]\Delta_{ij} \Big)\epsilon \nn\\
\delta \phi^i &=& i \bar \epsilon \gamma_5\Delta_i \Psi, \nn\\
\delta A_\nu &=& 
  i e^{-\sigma}\, G_{\nu\mu}\,\bar \epsilon\tilde\gamma^\mu \Psi
\eea
where 
$\tilde\Sigma^{\mu\nu} = \frac i4\,[\tilde\gamma^\mu,\tilde\gamma^\nu],$
$\Delta_{ij} = \frac 12 [\Delta_i,\Delta_j]$,
and $\Gamma_\mu = \gamma_\mu, \, \Gamma_{3+i} =  \Delta_i \gamma_{5}$
where $\Delta_i$ generates the $SO(6)$ Clifford algebra.
The constant $e^\sigma$ factors can be absorbed
by rescaling the fields.
This indeed amounts to the $\cN=4$ supersymmetry 
on $\R^4_\theta$.
In the case of general NC backgrounds,
the SUSY transformation will also act on the metric $G_{\mu\nu}(x)$.
This can be viewed as a supersymmetric form of emergent gravity, 
which should be worked out elsewhere.

\subsection{Symmetries and conservation laws}
\label{sec:symm}

The matrix model is invariant under the gauge 
symmetry \eq{gauge} as well as 
a global rotational and translational
symmetry. The gauge symmetry has been identified 
geometrically in section
\ref{sec:dof} as invariance under symplectomorphisms.
The $SO(D-1,1)$ resp. $SO(D)$
rotation invariance is realized as $X^a \to \Lambda^a_b\, X^b$
supplemented by suitable action on the spinors, and
translations act as
\be
X^a \to X^a + c^a\one, \qquad c^a \in \R .
\label{translations}
\ee
Together these form the inhomogeneous Euclidean group i.e.
the Poincar\'e group. As usual,
these global symmetries should lead to conservation laws.

%
%

Now consider the semi-classical case.
Taking advantage of this rotation and translation symmetry,
one can choose for any given point $p \in \cM$ adapted 
coordinates where the brane is tangential to the plane spanned by the 
first $2n$ components, i.e. $\partial_\mu \phi^i|_p =0$. 
Then the embedding metric satisfies
\be
\d g_{\mu\nu}|_p = \partial_\sigma  \d g_{\mu\nu}|_p =0 .
\label{normal-embedd-coords}
\ee
We denote such coordinates as 
``Normal Embedding Coordinates'' NEC
from now on; they are preferred matrix coordinates $x^\mu\sim X^\mu$.
This simplifies 
the analysis of general branes, allowing to reduce
many considerations to the case of trivially embedded branes. 
For example, it is now easy to see that the matrix model action 
for the $U(1)$ fields
can be written as in \eq{S-semiclassical-general}
in terms of 
\be
4\eta(x) = \{x^a,x^b\}\{x^{a'},x^{b'}\}\eta_{aa'} \eta_{bb'}
= e^\sigma\, G^{\mu\nu}(x) g_{\mu\nu}(x).
\ee
This is so because $\eta(x)$ can be viewed as a $SO(D)$ scalar, which 
in normal embedding coordinates reduces to 
$e^\sigma\,G^{\mu\nu}(x) \eta_{\mu\nu}$. The covariant 
version is as above.

If the effective metric  coincides with the embedding
metric $G_{\mu\nu} = g_{\mu\nu}$
as discussed below, then these NEC are 
``free-falling'' Riemannian normal coordinates in GR, 
i.e. $G_{\mu\nu}|_p = \eta_{\mu\nu}$ and 
$(\Gamma_{\mu\nu}^\rho)|_p = 0$. Local Lorentz invariance is 
then manifest in these coordinates. 
However, an important distinction with GR becomes 
obvious: 
the extrinsic curvature of the brane $\cM \subset \R^D$ 
does play a role in the matrix model
(at least in the ``harmonic branch'' discussed below), 
due to the brane tension. 
For example, even though a cylinder embedded in $\R^D$ is 
intrinsically flat, it is not a minimal surface.

\subsubsection{Tangential conservation law}

Motivated by the translational symmetry, one can consider the following
"$x$-dependent tangential" variation
\be
X^a \to X^a + \d X^a, \qquad
\d X^a = \{X^b,[X^a,\varepsilon_b(X)]\}
\label{dX-local}
\ee
which defines a measure-preserving transformation on the space of matrices.
The corresponding Dyson-Schwinger equations 
have the form of a conservation law 
\be
[X^a,T^{a'c}] \eta_{aa'} =0 .
\label{e-m-cons}
\ee
This provides useful non-perturbative 
information which is not restricted e.g. to 
BPS sectors. The bosonic action \eq{YM-action-1}  leads to the
matrix - ``energy-momentum tensor'' 
\cite{Steinacker:2008ri,AbouZeid:2001up,Das:2002jd,Okawa:2001if}
\be
T^{ab} =  [X^a,X^c][X^b,X^{c'}]\eta_{cc'} 
  + [X^b,X^c][X^a,X^{c'}] \eta_{cc'}
  - \frac 12 \eta^{ab} [X^d,X^c][X^{d'},X^{c'}] \eta_{dd'} \eta_{cc'} .
\label{T-ab} 
\ee
It is easy to verify this using
the matrix e.o.m., but \eq{e-m-cons} is more useful because it 
involves only tangential tensors on $\cM$. Indeed the semi-classical 
geometrical limit of $T^{ab}$ is 
given by 
\be
T^{ab} \,\sim\,  -2 G^{\mu\nu}(x) \del_\mu x^a \del_\nu x^b
+ 2 \eta^{ab} \eta(x) ,
\ee
which for $g_{\mu\nu} = G_{\mu\nu}$ is essentially the projector
on the normal bundle of $\cM \subset \R^D$.
As elaborated in  \cite{Steinacker:2008ya}, 
the conservation law \eq{e-m-cons} then leads to the following equivalent 
equations
\bea
\theta^{\nu\rho}\partial_\rho\eta
 &=& \theta^{\mu\eta}\partial_\eta(e^\sigma G^{\nu\rho} g_{\mu\rho})  
\label{cons-tang-explicit-1} \\
\partial_\eta(\rho e^\sigma\hat\theta^{\eta\nu})
&=& \rho\,\theta^{\rho\nu}\partial_\rho\eta
\label{conservation-general-1} \\
\nabla_\mu (\hat \theta^{\mu\nu})
&=& e^{-\sigma}\theta^{\mu\nu}\partial_\mu\eta 
\label{conservation-general} \nn\\
\nabla^\mu (e^{\sigma} \theta^{-1}_{\mu\nu})   
&=& e^{-\sigma} G_{\nu\nu'}\theta^{\nu'\mu}\partial_\mu\eta .
\label{conservation-general-2} 
\eea
where $\hat \theta^{\nu\eta}$ is 
anti-symmetric and defined in \eq{theta-hat}.
The last form coincides precisely with
the e.o.m \eq{theta-covar-id-text},
which therefore can be trusted at the quantum level, and it
holds more generally e.g. in the presence of a mass term or a 
more general potential. 
Note that the $D$ relations \eq{e-m-cons} reduce to $2n$ tangential ones,
while the "transversal" components give no new information.

This matrix conservation law also applies to the
nonabelian $SU(n)$ components. Then \eq{e-m-cons} gives precisely the 
Yang-Mills equations of motion for $SU(n)$ gauge fields
coupled to gravity obtained from \eq{action-expanded-2}. 
This seems to be the best
way to establish the action for nonabelian 
gauge fields \cite{Steinacker:2008ya}.

\section{Special geometries and perturbations}

The most obvious solution of the above covariant
equations of motion is of course $\R^{2n}$, corresponding
to the semi-classical limit $x^a = (x^\mu,0)$ 
of the Moyal-Weyl quantum plane $\R^{2n}_\theta$. 
Its effective metric $\bar G^{\mu\nu}$ is flat. 
In the supersymmetric case, these are 
BPS solutions \cite{Ishibashi:1996xs} and therefore protected from 
quantum corrections. 
To understand other, less trivial solutions,  
it is illuminating to introduce the following tensor
\be
\cJ^{\eta}_\g = e^{-\sigma/2}\, \theta^{\eta\g'} g_{\g' \g}
 = - e^{\sigma/2}\,  G^{\eta \g'} \theta^{-1}_{\g' \g} .
\ee
Then the effective metric can be written as
\be
G^{\mu\nu} 
= \cJ^{\mu}_\rho\, \cJ^{\nu}_{\rho'}\, g^{\rho\rho'}
= - (\cJ^2)^{\mu}_\rho\, g^{\rho\nu},
\ee
which is the reason for choosing the above normalization. 
$\cJ^{\rho}_\nu$ satisfies the following properties
\bea 
(\cJ^2)^{\mu}_\rho &=& - G^{\mu \nu} g_{\nu \rho}\nn\\
tr \cJ^2 &=& - 4 e^{-\sigma} \eta
\eea
due to the anti-symmetry of $\theta^{-1}_{\mu\nu}$. 
In $2n=4$ dimensions, we have the additional relations \cite{Steinacker:2008ya}
\bea
\det \cJ &=& 1  \nn\\
\cJ^2 + \cJ^{-2} &=& \frac 12 \tr \cJ^2\,\one 
\label{quartic-relation}
\eea
using \eq{sigma-rho-relation}. 
We can now  extend the discussion in section \ref{sec:euclid-mink}
on (anti)self-dual symplectic structures and Euclidean versus
Minkowski signatures to the general case.

\subsection{Self-duality and $G_{\mu\nu} = g_{\mu\nu}$.}
\label{sec:self-dual}

We focus on the $2n=4$ -- dimensional case.
Since everything is now formulated in tensorial language,
the same arguments
of section \ref{sec:euclid-mink}
apply at any given point $p \in \cM$.
Assume first that $\cM$ is Euclidean.
Then we can diagonalize the embedding metric at that point
$g_{\mu\nu}|_p = \d_{\mu\nu}$,
and bring the Poisson tensor into canonical form
\be
\omega = \theta^{-1}\,(\a\, dx^0 dx^3 \pm \a^{-1} dx^1 dx^2)
\label{omega-standard-E}
\ee
\eq{theta-standard-general-E} at $p \in \cM$
using a suitable $SO(4)$ rotation. 
This leads to\footnote{Of course this 
holds only at $p$ and does {\em not} extend into a local neighborhood
of $p\in \cM$.}
\be
G^{\mu\nu} 
= \diag(\a^2,\a^{-2},\a^{-2},\a^2) \qquad\mbox{at}\quad p \in \cM .
\ee
Noting that $\star (dx^0 dx^3) = dx^1 dx^2$
where $\star$ denotes the Hodge star
defined by $\varepsilon^{\mu\nu\rho\sigma}$ and 
$g_{\mu\nu}$ on $\cM^4$,
it  follows again that the corresponding symplectic form
is (anti-) self-dual ((A)SD) if and only if
\be
\star \omega = \pm \omega 
\quad \Leftrightarrow \quad
\a = 1 \,\,\,\mbox{resp.} \,\, e^{-\sigma}\eta = 1 
\quad \Leftrightarrow \quad G_{\mu\nu} = g_{\mu\nu} 
\quad \Leftrightarrow \quad \cJ^2 = -1 .
\label{selfdual-E}
\ee
These are tensorial statements. The 
second statement follows from 
\be
e^{-\sigma}\eta = \frac 12(\a^2+\a^{-2}) \,\, \geq \,\, 1 .
\label{eta-sigma-alpha}
\ee
Note that $\cJ^2 = -1$ means that $\cJ$ defines an almost-complex
structure, and $\cM$ becomes an almost-K\"ahler manifold.
It is easy to check that then \eq{theta-covar-id-text} reduces to
$\nabla^\mu \theta^{-1}_{\mu\nu} = 0$, 
which follows from $\star \omega = \pm \omega$  resp.
$d\star\omega = \pm d\omega = 0$. 

In the Minkowski case, completely analogous statements apply
once we adopt complexified Poisson structures with
imaginary time components,
having in mind a Wick rotation\footnote{A reader not comfortable  
with such a complexified $\theta^{\mu\nu}$ may prefer to
stick with the Euclidean case and 
postpone the issue of Wick rotation.} $x^0 \to it$ as discussed 
in section \ref{sec:euclid-mink}.
If we accordingly define $\varepsilon^{0123} = i$, then
\be
\star \omega = \pm \omega 
\quad \Leftrightarrow \quad
\a = 1 \,\,\, \mbox{resp.} \,\, e^{-\sigma}\eta = 1
\quad \Leftrightarrow \quad G_{\mu\nu} = g_{\mu\nu} 
\quad \Leftrightarrow \quad \cJ^2 = -1
\label{selfdual-M}
\ee
where 
\be
\omega = \theta^{-1}\,(i \a\, dx^0 dx^3 \pm \a^{-1} dx^1 dx^2) .
\label{omega-standard-M}
\ee
Notice that 
$\cJ^2 = -1$ as in the Euclidean case determines the reality structure of $\theta^{\mu\nu}$,
and is replaced by the quartic equation \eq{quartic-relation}
in the general case $g_{\mu\nu} \neq G_{\mu\nu}$.
This ensures that the physical metric 
\be
G^{\mu\nu} 
= \diag(-\a^2,\a^{-2},\a^{-2},\a^2) \qquad\mbox{at}\quad p \in \cM
\label{G-M-physical}
\ee
is always real. 

It is important to note that  
self-dual Poisson structures always solve the 
equation of motion \eq{theta-covar-id-text}, which reduce to
$\nabla^\mu \theta^{-1}_{\mu\nu} = 0$ for $e^{\sigma}=\eta$.
Such (A)SD closed 
2-forms $\theta^{-1}$ with constant asymptotics 
$\theta^{-1}_{\mu \nu}(x) \to \obar \theta^{-1}_{\mu \nu}$
as $x\to\infty$ always exist on (suitable) 
asymptotically flat spaces. 
This can be seen by interpreting 
$\theta^{-1} $ as 
sourceless electromagnetic field with 
constant field strength at infinity:
we only have to solve 
$d\star F = 0, \,\, F = dA$ with constant asymptotics 
$F \to \obar F$ as $r \to\infty$, and 
define $\theta^{-1} $ to be
the (A)SD component of $F$.
Therefore
\be
G_{\mu\nu} = g_{\mu\nu} , \qquad \nabla^\mu \theta^{-1}_{\mu\nu} = 0,
\qquad \Box \phi^i = \Box x^\mu = 0
\label{SD-solution}
\ee
provides a particularly interesting and 
transparent class of solutions without matter.
The bare action \eq{S-semiclassical-general} 
then has the form of a brane tension
using \eq{selfdual-E},
\be
S \,\,=\,\, \frac{\Lambda_0^4\,}{4 g^2}\, \int d^4 x\,\frac{\eta}{\theta^2} 
\,\,=\,\, \frac{\Lambda_0^4\,}{4 g^2}\,\int d^4 x\,\sqrt{|g|}\, .
\label{S-gG-tension}
\ee
This has the same form as the
one-loop vacuum energy, and the two may approximately cancel
provided $\L_0$ coincides with some effective cutoff.
In that case, the total effective brane tension may indeed be small. 
However, even though minimal surfaces 
are in general not BPS, they are protected  
and in fact stabilized by quantum corrections.

In the remainder of this review, we will mostly assume  that 
$\omega$ is (anti-) selfdual to a very good approximation, i.e.
 $G_{\mu\nu} = g_{\mu\nu}$. 
Most of the known solutions of matrix models
(possibly with additional quadratic and cubic terms) 
such as  fuzzy $\C P^2$ etc. 
indeed satisfy this condition. 
A more in-depth study of this class of geometries is 
given in \cite{curvature-BS}.
However, we now study small perturbations of this (A)SD case.

\subsection{Metric perturbations from $U(1)$ gauge fields}
\label{sec:linearized-U(1)}

Using the above results, the
most general geometry of the matrix model 
can be viewed as perturbation of an (A)SD background
$\bar G_{\mu\nu} = g_{\mu\nu}$
through ``would-be'' $U(1)$ gauge fields
$A_\mu$ with  field strength 
$F_{\mu\nu}$, leading 
to a metric perturbation 
\be
G_{\mu\nu} = \bar G_{\mu\nu} + h_{\mu\nu}
\label{linearized-h}
\ee
elaborated below.
These $U(1)$ gauge fields 
result from tangential variations of the matrices, while transversal fluctuations
of the matrices lead to variations of the embedding metric $g_{\mu\nu}$.
This appears to be the most useful parametrization.

Hence consider some $\cM\subset \R^D$ with embedding normal 
coordinates at $p\in\cM$ 
such that $g_{\mu\nu} = \diag(\pm 1,1,1,1)$ at $p\in\cM$. 
We can assume that the
Poisson tensor $\bar\theta^{\mu\nu}$ has the form 
\eq{omega-standard-M} resp. \eq{omega-standard-E},
leading to an effective metric 
\be
\bar G_{\mu\nu} = \diag(\pm\a^2,\a^{-2},\a^{-2},\a^2) 
\ee
at $p\in\cM$ as in \eq{G-MW-E},
which in the case of (A)SD Poisson structure 
satisfies $\bar G_{\mu\nu} = g_{\mu\nu}$. 
Now consider small fluctuations of the
tangential matrices 
$X^\mu$ around such a configuration. They can be
parametrized as 
\be
X^\mu = \bar X^{\mu} -\bar\theta^{\mu\nu} A_\nu(x) \, ,
\ee
and the $A_\nu$ were interpreted  in 
section \ref{sec:moyal-weyl-gauge}
as $U(1)$ gauge fields. On the other hand, we saw in 
section \ref{sec:dof}
that they can be interpreted as perturbation of the effective 
Poisson tensor \eq{theta-trafo} on $\cM$, 
\be 
\theta^{-1}_{\mu\nu} = \bar \theta^{-1}_{\mu\nu} + F_{\mu\nu}
\ee
where
$F_{\mu\nu} = \partial_\mu A_\nu - \partial_\mu A_\nu$. 
This in turn leads to a perturbation of the metric \eq{linearized-h} on $\cM$,
with
\bea
h_{\mu\nu} 
&=& \theta\,\Big((\bar \cJ^{-1})^{\mu'}_{\nu} F_{\mu'\mu}
 + (\bar \cJ^{-1})^{\nu'}_{\mu} F_{\nu'\nu}
 - \frac 12 ((\bar \cJ^{-1})^{\rho}_{\nu} F_{\rho\sigma}\bar G^{\nu\s})\bar G_{\mu\nu}\Big)
\label{hmunu-general}
\eea
where $\rho = |\theta^{-1}_{\mu\nu}|^{1/2}$ and
\be
(\bar \cJ^{-1})^{\g}_\mu
=  g^{\g\eta}\, \theta\,\bar\theta^{-1}_{\eta\mu}\,
= \,\left(\begin{array}{cccc} 0 & 0 & 0 & -\a^{-1} \\
                                0 & 0 & -\a & 0 \\
                                0 & \a & 0 & 0  \\
                                \a^{-1} & 0 & 0 & 0 \end{array}\right)\, .
\ee
To be specific we
assume the Euclidean case here, where
$\bar\theta^{\mu\nu}$ has the canonical form \eq{G-MW-E}.
As a check, observe that $h = \bar G^{\mu\nu} h_{\mu\nu}=0$
consistent with \eq{G-g-4D}.
This interpretation of the noncommutative 
$U(1)$ gauge fields
in terms of a metric perturbation
was first given by Rivelles \cite{Rivelles:2002ez}. He observed that
if the gauge fields 
solve the Maxwell equations, 
then the $h_{\mu\nu}$ are actually
gravitational waves resp. gravitons. 
To see this, consider the case of $\cM = \R^4$, so that 
the linearized Ricci tensor is given by 
\be
R^{(1)}_{\mu\g}  = 
\partial^\rho \partial_{(\mu}  h_{\g)\rho} - \frac12
\partial_\mu\partial_\g  h - \frac12 \partial^\d\partial_\d h_{\mu\g} .
\ee
Now recall the equations of motion  \eq{gamma-mu}
(equivalent to the Maxwell equations
for $F_{\mu\nu}$)
in the preferred (matrix) coordinates,
which together with \eq{tilde-Gamma} gives 
\bea
0 &=& \Gamma^\mu 
\sim \partial_\nu (\sqrt{|G_{\eta\sigma}|}\, G^{\nu\mu}) .
\eea
For the metric fluctuations this gives 
\be
\partial^\mu h_{\mu\nu} - \frac 12 \partial_\nu h =0 ,
\label{h-harmonic}
\ee
and the expression for the linearized Ricci tensor 
simplifies as
\be
R^{(1)}_{\mu\g}  = - \frac12 \partial^\d\partial_\d h_{\mu\g} .
\label{Ricci-harmonic}
\ee
Now  the Maxwell equations 
for the $U(1)$ fluctuations on $\R^4$
\be
\partial^\mu F_{\mu\nu} =0
\label{maxwell}
\ee
imply as usual
\be
\partial^\rho\partial_\rho F_{\mu\nu} =0 .
\label{box-F}
\ee
Together with \eq{Ricci-harmonic} and \eq{hmunu-general}, 
it follows that
these $U(1)$ metric fluctuations are Ricci-flat \cite{Rivelles:2002ez},
\be
R_{\mu\nu}[\eta_{\mu\nu} + h_{\mu\nu}]  =0 .
\label{ricci-flat}
\ee
This can  also be seen as a consequence of 
$\Box_{G} x^\mu =0$. Once we establish
that the corresponding Riemann tensor is 
indeed non-trivial, these are genuine gravitational waves\footnote{It was 
argued in \cite{Rivelles:2002ez} that these metric perturbations
generally correspond to pp waves.}
(in harmonic gauge \eq{h-harmonic}),
parametrized through on-shell degrees of freedom of
$U(1)$ gauge fields.
This provides an interesting relation with GR.

To make this more explicit,
consider the case $\bar G = g$ i.e. $\a=1$. 
Using the usual parametrization of $F_{\mu\nu}$ in terms of $E_i$ and
$B_i$,  
the corresponding metric perturbation is 
\be
h_{\mu\nu} = \theta\,
    \left(\begin{array}{cccc} (B_3-E_3) & (B_2- E_2) & -(B_1-E_1) & 0 \\
                              (B_2- E_2) & - (B_3-E_3) & 0 &  (B_1- E_1) \\
                            -(B_1-E_1) & 0 & - (B_3-E_3) &  (B_2- E_2)  \\
                            0 & (B_1- E_1) &  (B_2- E_2) & (B_3-E_3)
                          \end{array}\right) .
\label{h-traceless-explicit-euclid}
\ee
This vanishes for self-dual fields $B_i = E_i$
as expected due to \eq{selfdual-E} (only for $\a= 1$!), 
and is non-trivial for the anti-self dual
fields as we verify below. 
In the Minkowski case, the corresponding result for
\be
\bar\theta^{\mu\nu} = \theta\,
\left(\begin{array}{cccc} 0 & 0 & 0 & -i \\
                                0 & 0 & -1 & 0 \\
                                0 & 1 & 0 & 0  \\
                                i & 0 & 0 & 0 \end{array}\right)
\quad\mbox{and} \quad 
F_{\mu\nu} = \left(\begin{array}{cccc} 0 & i E_1 & i E_2 & i E_3  \\
                                -i E_1 & 0 & B_3 & -B_2 \\
                                -i E_2 & -B_3 & 0 & B_1  \\
                                -i E_3 & B_2 & -B_1 & 0
                              \end{array}\right) \,
\ee
is
\be
h_{\mu\nu} = \theta\,
    \left(\begin{array}{cccc} - (E_3+B_3) & -i(B_2+ E_2) & i(B_1+E_1) & 0 \\
                              -i(B_2+ E_2) & - (B_3+E_3) & 0 &  B_1+ E_1 \\
                            i(B_1+E_1) & 0 & -(B_3+E_3) &  B_2+ E_2  \\
                            0 & B_1+ E_1 &  B_2+ E_2 & E_3+B_3 \end{array}\right),
\label{h-traceless-explicit}
\ee
which vanishes for\footnote{recall that we use 
$\varepsilon^{0123} = i$ in the Minkowski case.} $\star F = F$.
The imaginary $h_{0i}$ are misleading,
since the physical metric in real coordinates
is manifestly real \eq{G-M-physical}. The above form is written
in ``partially Wick rotated'' coordinates, and the real
physical form is seen in coordinates where $h_{0i} = 0$. This
is obtained after a transformation 
$x^i \to x^i + x^0 \xi^i(x) = x^i + i t\xi^i(x)$,
which amounts to the required local Lorentz boost 
to achieve the normal form \eq{omega-standard-M}.
In case of doubt, one can first work in the 
Euclidean framework and do a formal Wick rotation in the end.

Hence on a self-dual background $g_{\mu\nu} = G_{\mu\nu}$,
only the anti-selfdual perturbations $F_{\mu\nu}$
describes non-trivial gravitational
waves, while the self-dual perturbations do not lead
to any metric deformations 
at least at the linearized  level.
However they are non-trivial perturbations of $\theta^{\mu\nu}$. Hence they 
seem to be completely sterile fields which decouple
from everything else. This suggests that they should be integrated
over. That would in particular suppress Lorentz-violating effects 
due to $\theta^{\mu\nu}$, such as in the  term
$\int d^4 x\, R_{\mu\nu\rho\sigma}\theta^{\mu\nu}\theta^{\rho\sigma}$
found in \cite{Klammer:2009dj}.
On the other hand if $\theta^{\mu\nu}$ is taken to be real 
in the Minkowski case, then
one finds 2 independent physical gravitons as in GR.

In summary, configurations with 
$g_{\mu\nu} = G_{\mu\nu}$ can provide the required geometrical degrees of
freedom for full-fledged gravity 
assuming $D=10$ \cite{clarke}, while perturbations due to the
``would-be'' $U(1)$ gauge fields are {\em additional} graviton-like degrees
of freedom on such a background. 
On the other hand they can also be interpreted in terms of the scalar fields
$\sigma(x)$ and $\eta(x)$.
Clearly more work is required in order to disentangle and fully understand the
significance of these degrees of freedom.

\paragraph{Plane waves and linearized Riemann tensor.}

We verify here that the 
Riemann tensor of the above metric perturbations is 
non-trivial.
Consider in particular plane waves
\be
A_\mu(x) = A_\mu e^{ikx} 
\ee
such that $F_{\mu\nu} = i(k_\mu A_\nu - k_\nu A_\mu) \neq 0$.
Then
\bea
h_{\mu\nu} &=& i\theta (\tilde k_{\mu}A_{\nu}+\tilde k_{\nu}A_{\mu})
 - i\theta (k_{\mu}\tilde A_{\nu} + k_{\nu}\tilde A_{\mu}) 
+  i\bar G_{\mu\nu}\,(\bar\theta^{\rho\sigma}k_{\rho}A_{\sigma})
\label{h-abdown}
\eea
where
\be
\tilde k_\mu = (\cJ^{-1})^\rho_\mu\, k_\rho .
\ee
Observe that the term 
$i (k_{\mu}\tilde A_{\nu} + k_{\nu}\tilde A_{\mu}) $
has the form of a diffeomorphism generated by $\tilde A_{\nu}$, and
therefore does not contribute to the Riemann tensor. 
Using \eq{h-abdown} for the metric fluctuations gives
\bea
R^{(1)}_{\sigma\mu\nu\g}
&=& - \frac i2\theta \Big(
( k_\mu \tilde k_{\s} - \tilde k_{\mu} k_\s)
(k_\nu A_{\g} - A_{\nu}k_\g)
+ (k_\mu A_{\s}  - k_\s A_{\mu}) 
(k_\nu\tilde k_{\g} - \tilde k_{\nu}k_\g)\Big) \nn\\
&&  - \frac i2\,( k_\nu k_\mu  \bar G_{\s\g} 
  - k_\nu k_\s  \bar G_{\mu\g} 
- k_\g k_\mu  \bar G_{\s\nu} 
+ k_\g k_\s  \bar G_{\mu\nu}) 
(\bar\theta^{\rho\sigma}k_{\rho}A_{\sigma}) 
\,\, \neq 0 .
\label{Riemann-first}
\eea
Thus the $U(1)$ metric fluctuations correspond to 
propagating gravitational waves on  $\R^4$ with
2 independent degrees of freedom, 
i.e. physical gravitons in harmonic gauge\footnote{the restriction 
to (A)SD field strength is not seen here because
there are no (A)SD plane waves.}.

\paragraph{Linearized coupling to matter.}

Here we briefly 
discuss the coupling of these ``would-be'' $U(1)$
degrees of freedom to matter, corresponding to perturbations of
$G_{\mu\nu} = g_{\mu\nu}$.
The coupling of $ h_{\mu\nu} = G_{\mu\nu} - g_{\mu\nu} = \d G_{\mu\nu}$
is given as usual by
\be
\d S_{\rm matter} \approx 8\pi\int d^4 x \sqrt{G}\, \d G_{\mu\nu} T^{\mu\nu}
\ee
where $T_{\mu\nu}$ is the energy-momentum tensor
of matter. 
We will focus on the case of static matter distributions with
$T_{00} = \rho(x), \,T_{ij} = T_{0i} =0$. 
Then the linearized metric
perturbation \eq{linearized-h} couples as
\bea
\d S_{\rm matter}  &\approx&  8\pi\int d^4 x \, h_{\mu\nu} T^{\mu\nu}
=  8\pi\int d^4 x \, h_{00}\, \rho(x) \nn\\
&\equiv& \int d^4 x \, (\vec E \cdot \vec m_E + \vec B \cdot \vec m_B)
\label{matter-g-lin}
\eea
in the Minkowski case, replacing 
$(-E_3-B_3) \to (B_3-E_3)$ in the Euclidean case. 
Hence from the point of view of electrodynamics, matter behaves like
a dipole\footnote{This is strongly reminiscent of the picture of
\cite{SheikhJabbari:1999vm,Bigatti:1999iz} where
open strings act as dipoles
in the presence of a $B$-field background.} with dipole strength 
\be
\vec m_E = -8\pi \theta (0,0,\rho(x)) = \vec m_B ,
\ee
pointing in the direction determined by $\theta^{0i}$. 
The total dipole moment is given by the mass $M$,
\be
\vec M_E = -8\pi M \theta \vec e_3 = \vec M_B, 
\ee
Note again that only the anti-selfdual components 
of the would-be $U(1)$ gauge field couple to matter.
In particular, observe that the total 
effective dipole moment of e.g. a system of stars rotating each other is
time-independent and pointing in a 
constant direction, hence they can radiate only
quadrupole and higher multipole modes in $h_{\mu\nu}$.

Hence matter leads to a small localized dipole-like
deviation from $\theta^{\mu\nu}(x)$ being self-dual, leading to
$g_{\mu\nu} \neq G_{\mu\nu}$ near matter. However, this 
will be suppressed by 
powers of $\frac{1}{\Lambda_0}$ resp. $\frac{1}{\Lambda_{NC}}$, 
and is therefore presumably negligible for most purposes.
We will therefore ignore the would-be $U(1)$ gauge fields, 
and focus on geometries with $G_{\mu\nu} = g_{\mu\nu}$.

\section{Quantization}
\label{sec:quantization}

Perhaps the main motivation for the matrix model approach 
is that they provide a natural non-perturbative concept
of quantization, defined in terms of an integral over all matrices
\be
Z = \int dX^a d\Psi e^{-S[X,\Psi]}
\label{matrix-pathintegral}
\ee
and similarly for correlation functions. 
In particular, the IKKT model thereby promises to provide an accessible 
quantum theory of gravity (as well as the other fundamental interactions),
because it can be viewed equivalently as 
$\cN=4$ NC SYM on $\R^4_\theta$. This allows to circumvent many of the 
technical problems in quantizing GR. 
In the Euclidean case, it was shown that this type of integrals 
exists for finite $N$ (in $D\geq 3$ dimensions) 
\cite{Austing:2001pk,Krauth:1998yu}, even for purely bosonic models; 
see also \cite{Kazakov:1998ji} for
some non-perturbative results. 
Nevertheless, in order to  obtain a well-behaved quantum theory, 
it is probably necessary to restrict to the IKKT model 
with $\cN=4$ supersymmetry resp. $D=10$, or
closely related models; this will be explained below.
Remarkably, $D=10$ happens to provide precisely the degrees of freedom 
needed to realize the most general 4-dimensional geometries
through embeddings $\cM^4 \subset \R^{10}$ \cite{clarke}. 
Even though the mechanism is quite different, it is worth
recalling that single-matrix models have been used as a 
non-perturbative definition of 2D gravity \cite{DiFrancesco:1993nw}. 

In the following we study the quantization around some NC background
as discussed above, which amounts to 
a ``condensation'' of the matrices. 
This is a strong assumption, which would imply that the 
$U(\infty)$ gauge symmetry is spontaneously 
broken\footnote{gauge fields can thus 
be viewed as analogs of Nambu-Goldstone
bosons, where the gauge symmetry is realized non-linearly.
This is a beautiful insight provided by NC gauge theory.}.
However, many of our results should hold also under somewhat
weaker assumptions, notably 
$\langle X^\mu\rangle = \langle[X^\mu,X^\nu]\rangle = 0$ but 
$\langle[X^\mu,X^\nu] [X^\mu,X^\rho]\rangle \neq 0$;
more specifically, $\theta^{\mu\nu}$ might 
not have long-range order but short-range fluctuations
and should be averaged.
This may also be important for suppressing 
Lorentz-violating effects.

\subsection{UV/IR mixing in noncommutative gauge theory}

Consider first the matrix model from the point of view of 
NC gauge theory on $\R^4_\theta$ as in 
section \ref{sec:moyal-weyl-gauge}.
A general matrix resp. field on $\R^4_\theta$ can be expanded in
a basis of plane waves,
\be
\phi(X) = \int \frac{d^4 k}{(2\pi)^4}\, \phi_k \, e^{i k_\mu \bar X^\mu}
\quad \in\,\, \Mat(\infty,\C) \cong \R^4_\theta
\ee
where $\phi_k \in \C$ 
is now an {\em ordinary} function of $k \in \R^4$.
The free (quadratic) part of the action is then independent of
$\theta^{\mu\nu}$, but the interaction vertices 
acquire a nontrivial phase factor 
$e^{\frac i2 \sum_{i<j} k^i_{\mu}k^j_{\mu}\theta^{\mu\nu}}$
where $k^i_{\mu}$ denotes the incoming momenta.
The matrix integral \eq{matrix-pathintegral} then becomes an 
ordinary integral 
$\int d X^a = \int \Pi d \phi_k$, 
which can be evaluated perturbatively in terms of 
Gaussian integrals, similar to ordinary QFT.
This leads to the well-known Filk-type Feynman rules 
\cite{Filk:1996dm} for field theory on $\R^4_\theta$,
where planar diagrams coincide with their 
undeformed counterparts, while the non-planar diagrams 
involve oscillatory factors. 
All this can be made rigorous on fuzzy spaces, where the 
matrices are finite-dimensional.
The reason for recalling these steps is
to emphasize that these rules are consequences of the 
basic definition \eq{matrix-pathintegral}, there
really is no choice\footnote{The generalization to 
the Minkowski case is mathematically more subtle, 
but formally the same derivation applies and
preserves the basic oscillating
character of the vertices. This should be well-defined
at least in the case of 
(softly broken) $\cN=4$ SYM, consistent with the 
concept of Wick rotation discussed previously.} 
in the framework of matrix-models. 

In particular, the loop integrals are generally divergent
in spite of the existence of a fundamental
length scale $\L_{NC}$. The reason is that area rather than length
is quantized on $\R^4_\theta$.
This implies that a UV--divergence
in some direction $k_\mu \ll \L_{NC}$ is necessarily associated
with an IR effect in another direction; this is the essence of
UV/IR mixing \cite{Minwalla:1999px}.  
Assuming some cutoff $\L$ in the loops, one finds
new terms in the one-loop effective action 
{\em for the trace-$U(1)$ components}, which are 
divergent for external momentum $p \to 0$. 
In the example of scalar fields coupled
to external $U(1)$, they have the form 
\bea
\Gamma_{\Phi} 
&=& - \frac{g^2}2\frac{1}{16 \pi^2}\,\int \frac{d^4 p}{(2\pi)^4}\, \, 
 \Big(-\frac 1{6} F_{\mu\nu}F_{\mu'\nu'}(-p) 
 g^{\mu\mu'} g^{\nu\nu'}
 \,\log(\frac{\L^2}{\L_{\rm eff}^2}) \nn\\
&& + \frac 14 (\theta F(p)) (\theta F(-p))
 \Big(\L_{\rm eff}^4 - \frac 16 p\cdot p\, \L_{\rm eff}^2 
 +\frac{(p\cdot p)^2}{1800}\, 
(47-30\log({\textstyle\frac{p\cdot p}{\L_{\rm eff}^2}})) 
 \Big)\Big)  
\label{ind-action-gauge}
\eea
where
\bea
\L_{\rm eff}^2(p) &=& 
 {1 \over 1/\Lambda^2 + \frac 14 \frac{p^2}{\L_{NC}^4}} .
\label{lambda-eff}
\eea
The first term amounts to a renormalization of the coupling, but the 
second term is unusual and singular as $p \to 0$. 
These IR divergences become worse in higher loops,
and the models are probably pathological as they stand.
There is a notable exception given by the
$\cN=4$ SYM theory on $\R^4_\theta$, 
which is nothing but the IKKT matrix model for $D=10$.
This is (almost\footnote{there is also a particular $\cN=1$ 
model with that 
property \cite{Jack:2001cr}, 
which however has the same field content and is closely
related to the IKKT model. More general soft SUSY breaking terms
(i.e. quadratic and cubic terms in the matrix model) should also 
be admissible.})  the unique model which has no 
UV/IR mixing at one loop, and is arguably well-defined
and finite to any order in perturbation theory  
just like its commutative cousin \cite{Jack:2001cr}.

Now focus on the $\cN=4$ model with soft SUSY breaking terms,
leading to a physical cutoff $\L = \L_4$ given by the scale
of $\cN=4$ SUSY breaking. Then the above ``unusual'' term in
$\Gamma_{\Phi}$ still leads to physically unacceptable 
effects from the point of view of $U(1)$ gauge theory (e.g.
polarization-dependent dispersion relations for the photon 
\cite{Jaeckel:2005wt}), 
however they are perfectly sensible from the point of view
of emergent gravity: they amount to an induced gravitational action,
as explained below.

\subsection{Induced gravity}

We now explain how this phenomenon of UV/IR mixing 
in NC gauge theory can be understood in terms of
induced gravity. The ``strange'' new term  in the effective 
action are simply induced gravitational terms\footnote{it seems likely that the interpretation of 
UV/IR mixing in terms of closed string modes
put forward in \cite{Armoni:2001uw,Sarkar:2005jw} is another way of looking at the same thing, since these closed-string
modes are related to the embedding metric $g_{\mu\nu}$ of the brane.} such
as an Einstein-Hilbert term $R[G]$.

\paragraph{Induced gravitational action due to scalars}

Consider the quantization of a scalar field
$\varphi$ coupled to the matrix model as in 
\eq{covariant-action-scalar}. 
Upon integration, this leads to an effective action 
\be
e^{-\Gamma_{\varphi}} = \int d\varphi\, e^{-S[\varphi]} , \quad\mbox{where}\quad
 \Gamma_{\varphi} = \frac 12 \Tr \log \Delta_{G} \, .
\ee
A standard argument using the heat kernel expansion
of $\Delta_{G}$ gives \cite{Gilkey:1995mj} 
\be
\Gamma_{\varphi} = \frac 1{16\pi^2}\, \int d^4 x \,\( -2\L^4 
- \frac 16 R[G]\, \L^2 + O(\log \L) \)\, .
\label{S-oneloop-scalar}
\ee
which is also the one-loop induced gravitational action due to 
a scalar field in the matrix model 
on a generally embedded brane $\cM^{4} \subset \R^D$ 
\cite{Grosse:2008xr}.
This is essentially the mechanism of induced gravity 
\cite{Sakharov:1967pk}.
It was shown in \cite{Grosse:2008xr} that this reproduces the 
UV/IR mixing terms \eq{ind-action-gauge} 
using the identification \eq{hmunu-general},
in the semi-classical limit
\be
p \L < \L_{NC}^2
\label{p-L-semiclassical}
\ee
where $p$ is the external momentum.
This means that the phase factors 
$e^{\frac i2 p_\mu k_\nu \theta^{\mu\nu}}$ 
in the non-planar loop integral
due to the non-commutativity are small and 
thus well reproduced by the semi-classical Poisson 
structure incorporated in $G_{\mu\nu}$.

\paragraph{Induced gravitational action due to fermions}

The one-loop effective action due to integrating out 
a fermion in the matrix model is more complicated.
Due to the non-standard spin connection, one cannot
simply use the standard results for the usual Dirac
operator. 
The induced gravitational action was determined 
in \cite{Klammer:2009dj} 
using the general formulae for the  Seeley-de Witt
coefficients for fermions. This is 
still too complicated in the most general case; 
for a Dirac fermion 
on generally embedded branes $\cM^{4} \subset \R^D$ 
with $G_{\mu\nu} = g_{\mu\nu}$,
the induced gravitational action  
can be written as
\bea
\Gamma_{\Psi} &=& \frac{1}{4 \pi^2}\int d^{4}x \sqrt{|g|}
\Big(2 \Lambda^4 +
\Lambda^2\Big(
-\frac{1}{3}R[g] + \frac{1}{4}\del_\mu\sigma \del^\mu \sigma \nn\\
&&\quad
+\frac{1}{8}e^{-\sigma} R[g]_{\mu\nu\rho\sigma}
\theta^{\mu\nu} \theta^{\rho\sigma}
+\frac{1}{4}(\Box_g x^a)(\Box_g x^b)\eta_{ab}\Big)
+\cO(\log \Lambda)
\Big).
\label{eff-action-final-2}
\eea
Here $k$ is the number of components of 
$D$-dimensional Dirac spinors. This formula generalizes
to the case of $2n$--dimensional branes.
The term $(\Box_g x^a)(\Box_g x^b)\eta_{ab}$
vanishes\footnote{This term can be expressed in terms of 
intrinsic quantities including $\theta^{\mu\nu}$.} 
for on-shell (tree-level) vacuum geometries, 
but may be non-trivial in general, reflecting
the embedding of $\cM^{4} \subset \R^D$.
A similar result was also obtained
for such on-shell (tree level) geometries
with $G_{\mu\nu} \neq g_{\mu\nu}$ \cite{Klammer:2009dj,Klammer:2008df}.
Again, it was verified that this reproduces the 
UV/IR mixing terms the fermionic loops  
in the semi-classical limit \eq{p-L-semiclassical}.
There is indeed a corresponding matrix model action 
 \cite{curvature-BS}, as given below.

\paragraph{Induced gravitational action due to gauge fields}

The induced gravitational action due to  integrating out 
gauge fields has not yet been computed directly for a 
general background. However, it can be obtained indirectly based on
supersymmetry: note that the
fermionic contribution to the 
one-loop effective action 
does not quite cancel the scalar contribution. 
Some UV/IR mixing may remain even in supersymmetric
gauge theory on $\R^4_\theta$, and
the UV/IR mixing is absent only in the case of $\cN=4$ supersymmetry
\cite{Jack:2001cr,Matusis:2000jf}.
This strongly suggests that the model should be finite 
above some scale $\Lambda$ to obtain a well-defined
quantum theory.
This is  realized by the 
IKKT model on a NC background, where
$\L$ is the scale of $\cN=4$ SUSY breaking above which  no divergent terms are induced.

Hence consider a matrix model 
 with $n_S$ scalar fields and $n_\Psi$ Dirac fermions.
The $\cN=4$ model has $n_S = 6$ and 
$n_\Psi = 2$, in addition to the $U(1)$ gauge field $A_\mu$.
Because this model is free of UV/IR mixing ,
it follows that
\be
\Gamma_A = - 2\, \Gamma_\Psi - 6\, \Gamma_\Phi \, 
\label{Gamma-A-explicit}
\ee
This matches the vacuum energy contribution
of 2 scalars \eq{S-oneloop-scalar}, but has a distinct
Einstein-Hilbert term\footnote{
This is the reason why even $\cN=2$ NC SUSY gauge theories 
do have UV/IR mixing.} as well as a  $R \theta\theta$
due to the would-be topological term \eq{FF-theta}.
Note that \eq{Gamma-A-explicit}
applies to both the $SU(n)$ as well as the $U(1)$ contributions
to induced gravity,
because all fields in the loop (abelian or nonabelian) couple in the
same way to external $U(1)$ legs. 

Thus  there will be no induced gravitational action
above the scale $\L$ of $\cN=4$ SUSY breaking. 
What happens below that 
scale depends on the specific breaking mechanism, which could be 
due to some Higgs effect, fermion condensation, or some other unknown
mechanism. The  $R \theta\theta$ may be problematic due to the 
explicit presence of $\theta^{\mu\nu}$ which breaks Lorentz
invariance. This term might disappear
due to fluctuations of $\theta^{\mu\nu}$, or
due to suitable SSB
such that the contributions from gauge fields and fermions cancel,
or by counterterms 
in the matrix model \eq{higher-terms}.
Then the effective gravitational action can take the 
desired Einstein-Hilbert form
\be
\Gamma_{grav} = \frac{1}{16\pi^2}\, \int d^4 x \,  \(\L_1^4 
+ R[G]\, \L_4^2 + O(\log \L) \)\, 
\label{induced-grav-optimist}
\ee
plus possibly some  additional terms as above.
Higher-order contributions in the curvature will be suppressed
by powers of $\frac 1{\L}$; note that
in induced gravity
there is no coupling constant of negative mass dimension
unlike in GR. Moreover, there are no unstable 
modes in the matrix model unlike in GR. 
Furthermore the strong RG running of the $U(1)$ sector is cut off at the 
NC scale. 
All this suggests that the quantization of 
gravity in this model should be well-defined and under control.
This would be jeopardized in the presence of 
higher-order terms such as 
\eq{higher-terms}, requiring and a more sophisticated 
RG-type analysis. 
In any case, the model certainly
provides an exciting new approach to quantum gravity.

\subsubsection{Higher-order terms in the matrix model}

As shown recently \cite{curvature-BS},
the above induced gravitational actions can be identified with higher-order 
terms in the matrix model. In particular,
\bea
\Tr
[[X^a,X^c],[X_c,X^b]][X_a,X_b] \,
&\sim& \int\! d^{4}x\sqrt{g} \(\frac{1}2 
\theta^{\m\eta}\theta^{\rho\a} R_{\mu\eta\rho\a} -2R + e^\sigma \del^{\mu}\sigma
\del_\mu\sigma\) \, \nn\\
\Tr\left([X^a,T^{bc}][X_a,T_{bc}]+T^{ab}\Box X_a \Box X_b\right) 
&\sim& \int\! d^{4}x\sqrt{g} \left((D-4)e^\s\Box_g e^\sigma-2e^{2\s}R\right)  
\label{higher-terms}
\eea
assuming $g_{\mu\nu} = G_{\mu\nu}$,
where $\Box X^a \equiv [X^c,[X_c,X^a]]$, and
 matrix indices are lowered with $\eta_{ab}$.
A matrix expression for $e^\s\Box_g e^\sigma$ term can also be found.
This could be used to cancel unwanted contributions in the induced action,
and to study systematically the quantization of gravity through 
purely bosonic matrix models.

\section{Macroscopic description: Harmonic branch and Einstein branch}

In this final section, we want to discuss the low-energy effective 
gravity theory arising from the $D=10$ matrix model, 
trying to minimize
any explicit reference to the detailed noncommutative origin.
We will derive equations of motion for 
gravity coupled to matter in the semi-classical 
limit, assuming that
$g_{\mu\nu} = G_{\mu\nu}$ 
with appropriate (anti-) self-dual $\theta^{\mu\nu}$.
This should be a very good approximation 
even in the presence of matter. 
Our starting point is the
semi-classical effective action of the matrix model 
\eq{S-gG-tension} together with the action for matter,
as well as the Einstein-Hilbert 
action $R[g]$ 
induced at one-loop
\be
S = \int d^4 x \sqrt{|g|}\, (\Lambda_4^2 R - 2\Lambda_1^4) + S_{\rm matter} ;
\label{basic-action}
\ee
possible additional terms are ignored for simplicity.
Now recall
\bea
\d \int\sqrt{|g|} &=& \frac 12 \int\sqrt{|g|} \, g^{\mu\nu} \d g_{\mu\nu},  \nn\\
\d \int\sqrt{|g|}\,  R &=& - \int\sqrt{|g|}\,\cG^{\mu\nu} \d g_{\mu\nu} \nn\\
\d S_{\rm matter} &=& 8\pi \int\sqrt{|g|}\,T^{\mu\nu} \d g_{\mu\nu} 
\eea
where $\cG^{\mu\nu} = R^{\mu\nu}- \frac 12 g^{\mu\nu} R$ 
is the Einstein tensor, 
and $T^{\mu\nu} $ is the energy-momentum tensor
for matter 
(recall that matter and fields couple to the effective metric 
essentially in the standard way). 
The crucial point is now that the fundamental
geometrical  degrees 
of freedom are not the $g_{\mu\nu}$, 
but the embedding fields $\phi^i$ as well as $X^\mu$ 
resp. $\theta^{\mu\nu}$.
The most general variation 
can thus be decomposed into variations of 
$g_{\mu\nu} = \eta_{\mu\nu} + \partial_\mu\phi^i \partial_\nu\phi^i$
and variations of $\theta^{\mu\nu}$. 
The e.o.m. for $\theta^{\mu\nu}$
are satisfied for $g_{\mu\nu}=G_{\mu\nu}$
to an excellent approximation as explained above. 
The variation of 
$S$ with respect to the fundamental fields $\phi^i$ 
can be written in matrix coordinates $x^\mu = X^\mu$ 
using $\d g_{\mu\nu} = \partial_\mu \phi^i \partial_\nu\d \phi^i  
+ \partial_\mu \d\phi^i \partial_\nu\phi^i $ as
\be
\d S = \int d^4 x \sqrt{|g|}\, 
\d g_{\mu\nu} \cH^{\mu\nu} 
= -2\int d^4 x \,\d \phi^i \partial_\mu(\sqrt{|g|}\,\cH^{\mu\nu})\partial_\nu\phi^i
\label{variation-cutoffs}
\ee
up to boundary terms, where 
\be
\cH^{\mu\nu} 
=  8\pi T^{\mu\nu}- \Lambda_4^2\cG^{\mu\nu} - \Lambda_1^4 g^{\mu\nu} .
\label{H-def}
\ee
This  leads to the equations of motion for $\phi^i$
\bea
\partial_\mu(\sqrt{|g|}\, \cH^{\mu\nu}\partial_\nu\phi^i) = 0 ,
\label{eom-general-H}
\eea
which using the identity $\nabla_\mu V^{\mu} \equiv
\frac 1{\sqrt{g}}\partial_\mu(\sqrt{g}\, V^{\mu})$
 can be written as
\bea
 \Lambda_1^4 \Box_g \phi^i
&=& (8\pi T^{\mu\nu} - \Lambda_4^2 \cG^{\mu\nu})\nabla_\mu\partial_\nu\phi^i
+ 8\pi (\nabla_\mu T^{\mu\nu})\partial_\nu\phi^i
\label{eom-general}
\eea
recalling  $\nabla_\mu\cG^{\mu\nu} =0$.
This equation has 2 types (``branches'') of solutions: 
\begin{enumerate}

\item
\underline{``Harmonic branch''}: 

The first class of solutions 
\be 
\partial_\mu(\sqrt{|g|}\, \cH^{\mu\nu}\partial_\nu\phi^i) = 0 , \qquad
\cH^{\mu\nu}  \neq 0
\label{harmonic-solution}
\ee
can be interpreted as solutions of a deformed Laplacian.
The prototype of such a solution 
is flat Minkowski space $\phi^i=0$.
Remarkably, this is a solution for arbitrarily large 
vacuum energy resp. $\Lambda_1>0$:
the term $\int d^4 x\sqrt{g}\, \Lambda_1^4$ is interpreted as
brane tension rather than as cosmological constant. 
More generally if the vacuum energy dominates the matter density,
\eq{harmonic-solution} reduces to $\Box_g \phi^i \approx 0$,
leading essentially to minimal surfaces.
As explained in more detail below, this
leads to the near-realistic cosmological 
solutions of FRW type \cite{Klammer:2009ku}, which are stable and largely 
insensitive to the detailed matter content. 
Remarkably, Newtonian gravity is also obtained 
due to the brane tension,
without even using the Einstein-Hilbert action.
In that scenario it is natural to assume that 
\be
\L_4 \ll \L_{\rm planck}
\ee
so that the induced gravitational action and in particular the
$R \theta\theta$ term \eq{eff-action-final-2}
are only small corrections and pose no
strong constraints.
Hence the harmonic branch may provide a solution of
the notorious cosmological constant problem, which is 
a strong motivation for more detailed studies.

Another attractive feature of this harmonic 
branch is that its quantization should be comparably
simple, as the embedding fields $\phi^i$ are governed
by a simple action with positive excitation spectrum.
On the other hand, it is questionable 
if the solar system precision tests can be met,
and more thorough studies are required.

\item
\underline{``Einstein branch''}: 

The above deformed harmonic equation \eq{harmonic-solution}
becomes void if $\cH_{\mu\nu} = 0$; then $\cM$ 
is in some sense a tensionless brane. This amounts precisely
to the Einstein equations 
\be 
\Lambda g^{\mu\nu} + \cG^{\mu\nu} = 8\pi G\, T^{\mu\nu} ,
\label{E-H-solution}
\ee
upon
identifying
\be
\frac {1}{G} =  \Lambda_4^2 \qquad
\frac {\Lambda}{G} = \Lambda_1^4 .
\label{E-H-indentification}
\ee
In particular, the Planck scale should be identified as
\be
\L_4 \approx \L_{\rm planck} ,
\ee
and the term $\int d^4 x\sqrt{g}\, \Lambda_4^4$
appears to be a cosmological constant.
Indeed using general embedding theorems \cite{clarke}, 
any solution of the E-H equations can be realized (locally)
using embeddings in $D\geq 10$. 
This provides a realization
of (induced) Einstein gravity through
matrix models.
Since the quantization is defined through 
the matrix model in a  background-independent way, 
certain problems of quantizing gravity 
such as unstable modes are  not expected to occur.
This should provide a 
quantum theory of gravity very similar to general relativity, 
along with all the other fields in the
IKKT model. Singularities should be resolved 
due to the quantum structure of
space-time. 
On the other hand, the usual cosmological constant 
and its fine-tuning problems appears to arise again.
However, the different parametrization of the geometry 
in terms of embedding rather than a fundamental metric might 
come to rescue: In the flat case, the first variation $\d \sqrt{g}$ due to 
transversal $\d\phi^i$ drops out (!), and the tangential 
variations are transmuted into variations of the Poisson tensor 
which also drop out as explained before. 
The same argument ``almost'' applies to near-flat geometries,
which should thus be at least meta-stable
even in the presence of large vacuum energy.
It remains to be seen whether the c.c. problem
can be solved in this way.

\end{enumerate}

Hybrid solutions also conceivable
where the rank of $\cH_{\mu\nu}$ is not maximal.
This is expected in the presence of extra dimensions
$\cM = M^4 \times K$, where $M^4$ could be a solution 
of the Einstein equations while $K$ 
could be e.g. a sphere (stabilized by a flux).
Indeed  the IKKT model provides just enough 
degrees of freedom for Einstein gravity on 
a 4-dimensional brane, 
but not for higher-dimensional ones.
In the presence of additional terms in the matrix model such as 
$Tr (X^a X^b X^c \varepsilon_{abc})$ or 
$Tr (X^a X^b \eta_{ab})$,
new types of solutions arise. Cubic terms
can be interpreted in terms of fluxes, and 
preserve the translational symmetry of the matrix model. 
The prototypical example of a solution in presence of a  
cubic term is the fuzzy sphere $S^2_N$, which is 
also a solution in the presence of a quadratic ``mass'' term.

The solar system precision tests 
clearly support the Einstein branch, and 
its realization in the present framework should be
studied in detail; e.g. the Schwarzschild geometry 
will be discussed in \cite{blaschke-steinacker}.
On the other hand, the harmonic branch is very interesting
because it leads to a stable cosmology which is remarkably
close to what we see, and to interesting 
long-distance modifications of gravity.  
We will therefore discuss the harmonic branch 
in the remainder of this paper. Although
at this early stage of investigation it does not appear to be 
fully realistic, some of the result might carry
over into a modified Einstein branch.

\section{Harmonic branch: gravity from branes with tension}

Based on the general results above,
we can start to explore the physical properties 
of gravity in this matrix model.
In this section some preliminary
and particular results for the ``harmonic branch''
of the model will
be discussed, which may or may not be physically
relevant. These solutions can be understood largely  
in terms of embedded branes $\cM\subset \R^{10}$ 
with tension coupled to matter. 
We will identify an intuitive and physically appealing
mechanism for (Newtonian, at least) gravity 
on the brane, which applies in a more general context.

\subsection{Cosmological solutions}
\label{sec:cosmology}

Consider first cosmological 
solutions of Friedmann-Robertson-Walker (FRW) type. 
These are characterized by spatially homogeneous and isotropic metrics 
\be
ds^2 = -dt^2 + a(t)^2 d\Sigma^2, 
\qquad d\Sigma^2 = \frac 1{1-k^2 r^2} dr^2 + r^2 d\Omega^2 
\label{FRW-0}
\ee
where $d\Sigma^2$ is the metric on a 3-dimensional
space with constant curvature, and $a(t)$ is the cosmological
``scale'' parameter. There are 3 possibilities: the space-like 3-manifold is flat $\R^3$ 
(denoted by $k=0$), a 3 sphere $S^3$ with $k=+1$, or hyperbolic space 
$H^3$ where $k=-1$.
In GR, such FRW solutions are obtained in the presence of a 
homogeneous matter distribution, which leads to the Friedmann
equations for $a(t)$.

In the present framework, we will look for solutions 
describing an embedding
$\cM^4 \subset \R^{10}$ with metric $G_{\mu\nu} = g_{\mu\nu}$
of FRW type. We know from  section \ref{sec:self-dual} 
that this is a {\em vacuum} 
solution of the matrix model 
provided $\cM^4 \subset \R^{10}$ is a minimal surface i.e. 
$\Box x^a =0$, equipped with a (anti-) self-dual nondegenerate
2-form $\theta^{-1}_{\mu\nu}$.
The presence of matter will in general modify these solutions, however
they are expected to be a good approximation 
provided the vacuum energy $\Lambda_1^4$ dominates the 
energy density $\rho$ due to matter. Thus we will assume that 
$\Lambda_1$ is at least of order $eV$, say, and increasing
$\Lambda_1$ will only stabilize the solutions. This is in striking
contrast with GR, where the vacuum energy resp. cosmological constant
must be very precisely tuned to $\L_1 \approx 2 meV$ according
to the $\L$CDM model. This leads naturally
to very reasonable solutions which are in remarkably good 
agreement with observations (leaving aside aspects related to the early
universe such as the CMB, which need to be re-analyzed 
in the present framework).

Generalizing the one-parameter solutions  in \cite{nielsen},
the following embeddings of FRW geometry through minimal surfaces
$\cM^4 \subset \R^{10}$
were found in \cite{Klammer:2009ku}:
\be
x^a(t,\chi,\theta,\varphi) = \(\begin{array}{c}
\cR(t) \(\begin{array}{l} \sinh(\chi)\sin\theta\cos\varphi \\
     \sinh(\chi)\sin\theta\sin\varphi  \\
     \sinh(\chi)\cos\theta \\
     \cosh(\chi) \end{array}\)\\
0 \\  x_{c}(t)\end{array}\) \in \R^{10} \nn
\label{cosm-embedding}
\ee
where
\be
\cR(t) = a(t)\,\(\begin{array}{l} \cos\psi(t) \\
\sin\psi(t) \end{array}\) 
\label{R-rotation}
\ee
and $\eta_{ab} =\diag(+,...+,-,-,+,+)$.
This leads to a FRW geometry with $k=-1$  
\be
ds^2 = -dt^2 + a(t)^2 d\Sigma^2,  \qquad
d\Sigma^2 = d\chi^2 + \sinh^2(\chi)d\Omega^2 
\label{FRW}
\ee
where $r = \sinh(\chi)$. This
is harmonic $\Box\, x^a =0$ provided
\bea
0 &=& \Box (\cR(t) \sinh(\chi)\cos\theta)  \nn\\
0 &=& \Box x_c .
\label{xdot-eq}
\eea
This leads to
\bea
3\frac 1{a}\,(\dot a^2-1) + \ddot a - \dot\psi^2 a &=& 0 \label{addot-eq}\\
5 \dot \psi \dot a  + \ddot \psi a &=& 0 \label{psidot-eq}\\
3\frac 1{a}\,\dot a\dot x_c + \ddot x_c &=& 0 .
\label{a-eom}
\eea
Similar solutions for 
$k=+1$ were also found which are however unrealistic, while
for $k=0$ only the one-parameter solution in \cite{nielsen}
is known.
The equations \eq{a-eom} can be integrated as follows:
\bea
(\dot a^2-1)a^6 + b^2 a^{-2} &=& d^2 = const  \nn\\
a^5 \dot \psi  &=& b\, = const >0  \nn\\
a^3 \dot{x_c} &=& d\, = const  ,
\label{cosm-firstorder}
\eea
leading to a 2-parameter family of solutions;
the case $b=0$ was obtained before in \cite{nielsen}.
For the Hubble parameter this implies
\bea
H^2 = \frac{\dot a^2}{a^2} &=& -b^2 a^{-10} + d^2 a^{-8} +\frac {1}{a^2} ,
\label{H-equation} \\
\frac{\ddot a}{a} &=& - 3 d^2 a^{-8} + 4 b^2 a^{-10} .
\label{addot}
\eea
For the early universe i.e. small $a$  and $b\neq 0$, 
this leads to a big bounce with minimal size 
$a_0 \sim b^{1/4}$. There is a transition determined by $d$
to a coasting universe at late times with
$\ddot a <0,\,\,\dot a \to 1$. 
The physics of the early universe in this model
is quite different from standard cosmology and 
requires a more detailed analysis including matter\footnote{
In particular the consistency with the CMB data 
cannot be reliably addressed at this point, 
see however \cite{BenoitLevy:2008ia} for the simplified case of 
an exact Milne universe. The above refined solutions 
yield a big bounce and 
an early phase with power-law acceleration \cite{Klammer:2009ku},
but the incorporation of matter is still missing.};
however,  it is a solid prediction that 
$a(t) \to t$ for late times, i.e. the solution approaches a 
Milne universe. This is in remarkable
good agreement with the basic observations
including the type Ia supernovae data \cite{BenoitLevy:2008ia}, 
which are usually interpreted in terms of an accelerating
universe.

\paragraph{Milne universe.}

The Milne universe is nothing but (a quarter of) 
flat Minkowski space $\R^4_+$, 
with flat metric
$d s^2 = -d\tau^2 + dr^2 + r^2 d\Omega^2$ 
written in terms of the  variables 
\be
\tau =  t \cosh(\chi), \,\, r =  t \sinh(\chi).
\ee
Then this metric takes the form of a FRW metric with 
$a(t) = t$ and $k=-1$,
\bea
d s^2 &=& -  d t^2 +  a(t)^2 (d\chi^2 +\sinh^2(\chi)d\Omega^2) .
\eea
This is the metric in the above solution for late times, 
which is expected to be valid
as long as the vacuum energy $\Lambda_1^4$ dominates the 
energy density due to matter. 
The solution is now easy to understand:
it is an almost-trivial embedding $\R^4_+\subset \R^{10}$ 
with a slight rotation in the early universe,
like a twisted rubber sheet.
Clearly such a solution is  stabilized by  large $\Lambda_1$.

To obtain a reasonable theory of gravity, 
it remains to be shown that 
matter such as stars and galaxies lead to appropriate local
perturbations of this cosmological solution, and reproduce
at least Newtonian gravity.
This indeed happens in this 
harmonic branch through an interesting mechanism, 
as we show next.

\subsection{Perturbations and
Newtonian gravity from brane tension} 
\label{sec:perturbations}

Now consider perturbations of the flat embedding 
$\R^4\subset \R^D$ through
the scalar fields $\phi^i$, leading to 
\be
g_{\mu\nu} = \eta_{\mu\nu} 
+ \partial_\mu\phi^i\partial_\nu\phi^j\, \d_{ij} 
\qquad \equiv \,\,\eta_{\mu\nu} + h_{\mu\nu} ,
\ee
keeping only terms linear in $h_{\mu\nu}$.
Clearly we must keep contributions to 2nd order in $\phi^i$
in order to get any non-trivial metric, hence
Newtonian gravity will arise through a non-linear mechanism. 
We assume that $G_{\mu\nu} = g_{\mu\nu}$ as discussed above,
and focus on {\em static} metrics $g_{\mu\nu}$, 
corresponding to static 
and somewhat localized matter distributions.
Thus consider the following localized excitation of the 
embedding 
\be
\phi^i(x,t) = g(x) e^{i\omega t} 
= g(x) \(\begin{array}{c}\cos(\omega t) \\ \sin(\omega
  t)\end{array}\), \qquad i=1,2
\label{harmonic-ansatz}
\ee 
with very small $\omega$. This
leads to the  metric 
\be
d s^2 = -(1- \omega^2 \, g^2) \, dt^2 + 
(\d_{ij} + \partial_i g \partial_j g) dx^i dx^j ,
\label{basic-metric-static}
\ee
which is static, because \eq{harmonic-ansatz} is
a standing wave rather than a traveling wave.

\subsubsection{Harmonic gravity bags}

Consider first vacuum excitations $\phi_0^i(x,t)$
for $\rho=0$. 
Neglecting corrections due to the induced gravity action,
the equation of motion is simply
\be
\Box \phi_0^i(x) = 0  
\ee
which for spherical waves reduces to
$\partial_r(r^2 g') + \omega^2 r^{2} g(r) = 0$.
The unique localized spherically symmetric solution which is
regular at the origin is 
\be
g_0(r) = g_0 \frac{\sin(\omega r)}{\omega r} , \qquad
\phi_0(x) = g_0(r)\, \(\begin{array}{c}\cos(\omega t) \\ \sin(\omega t)\end{array}\)
\label{phi-0}
\ee
with radial wavelength given by 
\be
L_\omega = \frac{2\pi}{\omega} .
\label{L-omega}
\ee
The effective metric \eq{basic-metric-static} is
\be
d s^2 = -(1- \omega^2 \, g(r)^2) \, dt^2 + (1+(g')^2) dr^2 
+ r^2 d\Omega^2 ,
\label{basic-metric-static-r}
\ee
which allows to read off the effective
gravitational potential $U_0$  seen by
a static test particle: 
\bea
g_{00} = -(1+2U_0), \qquad 
2U_0(r) &=& - \omega^2 \, g(r)^2 =  -\omega^2 g_0^2 
\Big(\frac{\sin(\omega r)}{\omega r}\Big)^2 , 
\eea
which satisfies
\be
U_0(r) \sim \left\{\begin{array}{ll}
- \frac{\omega^2}{r^2}, & r \to \infty  \\[1ex]
 -\frac 12 g_0^2 \omega^2 , & r \sim 0 
\end{array}\right.
\label{U-ac}
\ee
Thus $U(r)$ describes 
an attractive ``gravity bag' with size $L_\omega$, 
decreasing as $\frac 1{r^2}$ for $r>L_\omega$. 
Due to the attractive 
gravitational force, matter will tend to accumulate inside
these gravity bags. 
In particular, large clusters of matter such as galaxies will  
be embedded in such gravity bags.
The essential point is that the matter {\em within} such a 
gravity bag will experience Newtonian gravity,
due to a local deformation of the gravity bag. 

\subsubsection{Perturbed gravity bags, 
Newtonian gravity and Poisson equations}
\label{sec:Newton}

\begin{figure}
\begin{center} 
  \vspace{-0.2cm}
 \includegraphics[scale=0.45]{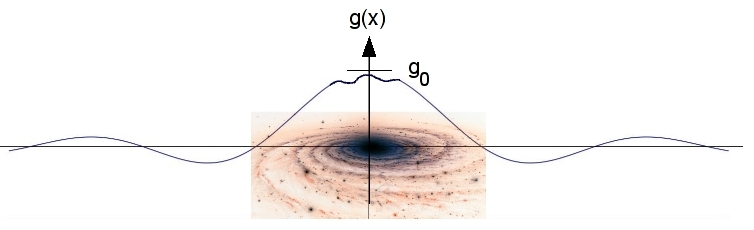}
\end{center}
\caption{sketch of embedding function $g(x)$ with short-scale perturbations and 
long-distance oscillations}
\label{fig:galaxy}
\end{figure}

To understand the mechanism,
consider first a spherically symmetric 
static mass density $\rho$
around the origin within the radius $r_M$. For $r>r_M$,
$\phi^i(x)$ is again a solution
of $\Box \phi^i =0$ and therefore must have the form
\bea
\phi^i &=& g(r) e^{i \omega t} , \qquad
g(r) =  g_0 \frac{\sin(\omega r+\d)}{\omega r} 
\quad   \sim \,\,  g_0 (\cos(\d) + \frac{\sin(\d)}{\omega r}) 
\label{g-r-phase}
\eea 
assuming $\omega r\ll 1$.
The phase shift $\d\ll 1$ is due to the presence of 
matter at the origin, and 
is the key for obtaining Newtonian gravity.
The effective metric \eq{basic-metric-static-r} becomes 
\be
g_{00} 
= -1 + g_0^2 \omega^2\cos(2 \d) + \frac{g_0^2 \omega\sin(2\d)}{r} 
\quad +O ((\frac{\d}r)^2) + O(r) 
\label{g00-simple} 
\ee
and similarly for $g_{rr}$; the correction terms will 
be discussed later.
This corresponds to a gravitational potential 
which for intermediate distances
\be 
\sin\d \ll \omega r \ll 1
\ee 
is well approximated by a $\frac 1r$ 
potential with a constant shift,
\be
U(r) = - \frac 12 \omega^2 g^2 
\,\,\approx\,\, U_0 -\frac{g_0^2
  \omega\sin(2\d)}{2r} ,
\qquad U_0 \approx  - \frac 12  \omega^2 g_0^2 .
\label{U-0-bag}
\ee
The phase shift $\d$ indeed turns out to be 
proportional to $M$, which gives Newtonian gravity.
To understand this more generally,
consider the e.o.m. \eq{eom-general} for 
$\phi^i$ coupled to matter,
\be
 \Box_\eta \phi^i =  \frac{8\pi }{\Lambda_1^4}\, \tilde T^{\mu\nu}
\partial_\mu\partial_\nu\phi^i , \qquad
\tilde T^{\mu\nu} \equiv T^{\mu\nu} - \frac{\Lambda_4^2}{8\pi } 
\cG^{\mu\nu}
\label{eom-general-linear}
\ee
replacing $\Box_g \approx \Box_\eta$ 
and using $\nabla_\mu T^{\mu\nu}=0$. 
Assuming 
$\rho(x) =\tilde T_{00} \geq 0$ and
$\tilde T_{ij} \approx 0$ in the presence of (non-relativistic)
matter\footnote{This is not evident, as the
simple embeddings below 
will generally {\em not} lead to $\cG^{ij} = 0$. 
It might be justified either for more 
sophisticated embeddings, or 
-- more interestingly -- if $\Lambda_4 \ll \Lambda_{\rm planck}$
which is very appealing as we will see.}, this becomes
\be
\L_1^4 \Box \phi^i = 8\pi \rho \partial_0^2  \phi^i .
\label{eom-matter-static-2}
\ee
The solution of this equation depends on the
amplitude of the background ``gravity bag'' $\phi_0(x)$
outside of the matter distribution. 
Since we are interested in the gravitational field
due to e.g. a star within a galaxy or some large-scale
cosmic mass distribution, we make the ansatz
\be
\phi^i(x,t) = g(x) e^{i\omega t} , \qquad
g(x) = g_0(x) + \d g(x)
\label{harmonic-ansatz-general}
\ee
where $|\d g| \ll g_0(x) \approx g_0(0)$ 
is varying on short scales
according to $\rho(x)$, while $g_0(x)$ is slowly 
varying at the scale $L_\omega$. 
Thus $g_0(x)$ reflects
the average mass distribution in the galaxy resp. 
a large cosmic structure, while $\d g(x)$ is 
a small local perturbation due to e.g. a single star,
as illustrated in figure \ref{fig:galaxy}.
The separation $g(x) = g_0(x) + \d g(x)$ corresponds to 
the splitting in \eq{g-r-phase}. Now it makes sense to
linearize in $\d g(x)$; this is the crucial step. 
Then \eq{eom-matter-static-2} gives
\be
(\Delta+\omega^2) g =  -8\pi\omega^2\frac{\rho}{\Lambda_1^4} g(x) .
\label{eom-embedd-1}
\ee
which for $g(x) = g_0(x) + \d g(x)$
with $\d g \ll g_0$ implies 
\be
(\Delta + \omega^2)\d g =  - 8\pi\omega^2\frac{\rho}{\Lambda_1^4} (g_0 + \d g)
\approx  - 8\pi\omega^2\frac{\rho}{\Lambda_1^4} \, g_0 .
\label{eom-embedd-2}
\ee
The gravitational potential is then given by
\be
U(x) = U_0  - \omega^2 g_0 \d g(x) \quad + \, O(\d g^2)
\label{U-newton-linfluct}
\ee
(cf. \eq{U-0-bag}), which under the above assumptions satisfies
\be
(\Delta + \omega^2) U(x) \approx
-  \omega^2 g_0 (\Delta+\omega^2) \d g(x) 
=  8\pi\omega^4 g_0^2 \frac{\rho(x)}{\Lambda_1^4}.
\ee
This is the desired Poisson equation
\be
\Delta U \approx  4\pi G\, (\rho(x) + \frac{\Lambda_1^4}{8\pi})
\,\, = \,\,  4\pi G\, \rho(x) - \Lambda_{\rm eff}
\label{poisson-eq-q}
\ee
where we identify
\bea
G &=& 2g_0^2\,\frac{\omega^4}{\Lambda_1^4} 
= - 4 U_0 \frac{\omega^2}{\Lambda_1^4} ,
\label{Newton-constant}\\
\Lambda_{\rm eff} &=& - \frac 12 G \Lambda_1^4 = 2 U_0 \omega^2,
\label{Lambda-eff}\\
U_0 &=& - \frac 12 g_0^2 \omega^2 .
\label{U-const}
\eea
Note that $G$ is naturally small 
since $|U_0| \leq 1$ and $L_\omega$ is large.
The correction due to $\Lambda_1^4$ 
can be identified as 
effective vacuum energy resp. 
apparent negative cosmological 
constant $\Lambda_{\rm eff}$.
If $\rho \gg \Lambda_1^4$
this reduces to the usual Poisson equation of Newtonian gravity,
$\Delta U = 4\pi G\, \rho$.
Corrections to the Newtonian approximations 
will be discussed below.

We conclude that localized matter $\rho$ inside 
a gravity bag 
is subject to  Newtonian gravity, with a dynamically
determined  gravitational ``constant'' $G$ 
given by \eq{Newton-constant}.
For example, two stars or planets would lead to 
local perturbations
\be
g(x) = g_0 + \d g_1(x) + \d g_2(x)
\ee
where $\d g_i$ are perturbations due to 
object $\rho_i$. They both see the same $g_0$ and $\omega$, 
thus the same gravitational constant $G$, and 
Newtonian gravity is recovered 
on scales shorter than $L_\omega$.
However at very long scales $L_\omega$ and 
near the border of the galaxy resp. the gravity bag, 
the effective gravitational
constant will vary.
The role of the 
``extrinsic curvature'' of the gravity bag 
and its brane tension will be clarified  
in section \ref{sec:local-bag}
using a different approach.

\paragraph{Gravitational (non--)constant $G$.}

Since $L_\omega, g_0$ 
and therefore $G$ are dynamical here,
the question arises why $G$ would be 
at least approximately 
the same in different parts of the universe. 
A possible explanation might be as follows:
Galaxies are typically 
parts of large structures such as (super)-clusters and filaments, 
which should provide the 
dominant contribution to the brane embedding and hence to 
$G$. Given the homogeneity of the CMB background, it 
is plausible that similar scales for $g_0(x)$ arise
on these dominant cosmic structures.
As a consistency check, one can show that $\omega$ and $g_0$ of
gravity bags propagating in a
flat Milne resp. Minkowski background 
remain essentially constant in time \cite{Steinacker:2009mp}.

Hence we obtain at least 
a crude approximation of the universe as we see it
at late times, with far less fine-tuning than in 
the standard model.
One may hope that a fully realistic picture
will arise in a more sophisticated treatment.

\subsection{Beyond Newtonian gravity}
\label{sec:corr-Newton}

Now consider spherically  symmetric mass distribution
$\rho(r)$ at the origin. 
A more careful analysis \cite{Steinacker:2009mp}
shows that 
the time-component of the effective metric 
is given by
\be
g_{00} = - \(1 +2 U_0 - 2 \frac{G M}{r} 
 + \frac 23 M G \,\omega^2 r - \frac 13 \Lambda_{\rm eff} r^2
+ (\frac{M G}r)^2 \frac 1{2U_0} \)\,\, + O(r^3) .
\label{metric-spherical}
\ee
We assume that the Newtonian potential due to  
$M$ is smaller
than the background potential $U_0$ due to the 
harmonic bag,
\be
 \frac{M G}{r} <  |2U_0| .
\label{outside-condition}
\ee
Then $O(M^2/r^2)$ term 
can be neglected, and the vacuum energy term
(as well as the Newtonian potential) dominates the 
linear term,  
$|\Lambda_{\rm eff} r^2| = 2 |U_0| \omega^2 r^2 \gg 
 \frac{M G}r \omega^2 r^2$.
Then $g_{00}$  has 
approximately the form of a Schwarzschild-de Sitter 
metric with apparent negative (!) cosmological
constant $\L_{\rm eff}$ and a constant shift \cite{Rindler:2006km},
\be 
g_{00} \approx - \Big(1 +2 U_0 - \frac{2G M}{r} 
 - \frac 13 \Lambda_{\rm eff} r^2 \Big)
\label{g00-approx}
\ee
assuming $r < L_\omega$. 
The Newtonian term  dominates the $\Lambda_{\rm eff}r^2$ term 
provided the vacuum energy 
$E_{\rm vac}(r) = \frac{4\pi r^3}{3} \Lambda_1^4$
is smaller than $M$.   
We then obtain Newtonian gravity with
potential \be U(r) \,\, \approx \,\, U_0 - \frac{G M}{r} .
\label{U-Newton-r} 
\ee
The radial part of the effective metric 
turns out to be 
\be
g_{rr} \approx 1+ \frac 13 \frac{2GM}{r} 
- \frac 19  \Lambda_{\rm eff} r^2\,\, 
 - \frac{GM}{r} \frac{1}{2\Lambda_{\rm eff} r^2} \frac{GM}{r}(1+\omega^2 r^2) 
 \, + O(r^3) .
\label{grr-simple-2}
\ee
The factor $\frac 13$ differs from general
relativity, and will be confirmed  
in section \ref{sec:local-bag} using a different approach.
This presents a challenge
for the solar system constraints; however
this might change in more sophisticated embeddings, and
a more complete analysis 
is required before a reliable judgment can be given.

The basic result is that 
Newtonian gravity arises at intermediate scales, with
significant long-distance modifications.
Notice that the precise form of the induced gravitational
action was never used up to now, rather
gravity arises through a deformation of the 
harmonic embedding  which couples to $T^{\mu\nu}$.
Thus the mechanism is quite different from GR.

\paragraph{Vacuum energy and cutoff.} 

\begin{figure}
  \vspace{-0.3cm}
\begin{center} 
\hspace*{0.2cm}
\includegraphics[scale=0.2]{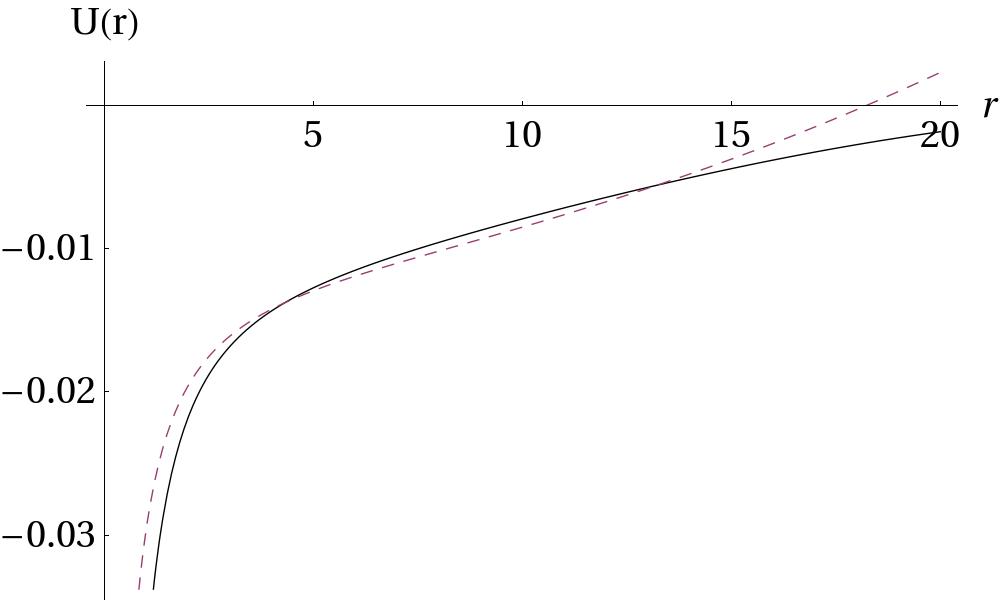}
\hspace*{-1cm}
\includegraphics[scale=0.2]{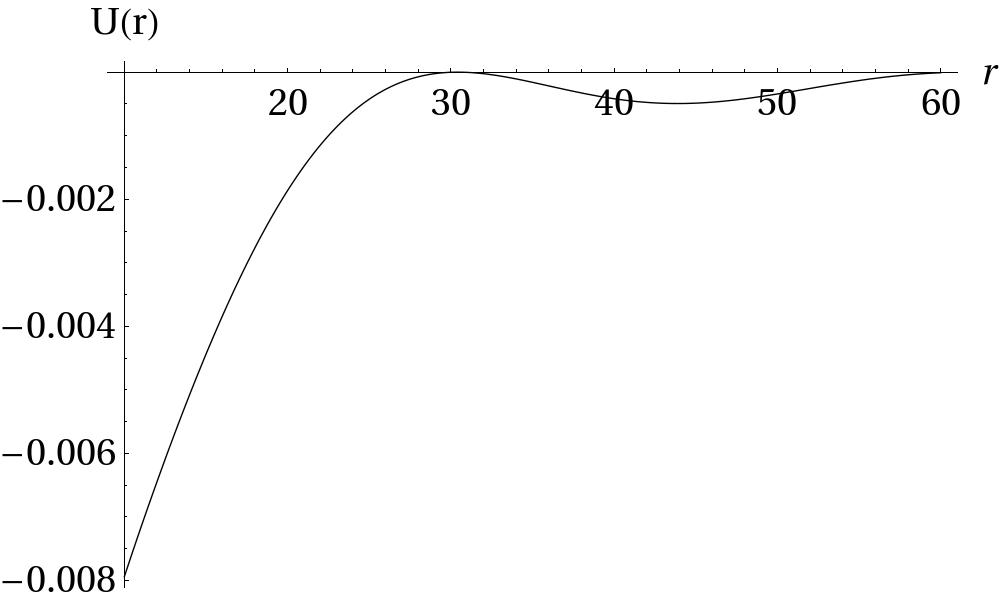}
\end{center}
\caption{Gravitational potential $U(x)$  compared with the
Schwarzschild-de  Sitter potential 
$U_0 - \frac{MG}{r} - \frac 16  \Lambda r^2$ 
(dashed line) and long-distance oscillations, for
$\omega = 0.1, \, g_0 = 1, \d = 0.1$.}
\label{fig:U}
\end{figure}

Inside the gravity bag, the
vacuum energy $\L_1^4 > 0$ contributes a positive energy 
density to the gravitational potential
\eq{poisson-eq-q} within the harmonic gravity bag,
leading to an additional gravitational binding.
However
for very large distances $r \geq L_\omega$, the harmonic behavior
\be
U(r) \sim - \frac 12 \omega^2 g_0^2 \frac{\sin^2(\omega r)}{r^2}
\ee
dominates and
leads to a screening of gravity, smoothly merging 
$g_{\mu\nu}(x) \to \eta_{\mu\nu}$ with the flat metric 
of the Milne-like cosmology 
as discussed in section \ref{sec:cosmology}.
Recall that cosmology does not lead to the usual 
stringent constraints on the vacuum energy here.
An illustrative plot of  $U(x) = -\frac 12 \omega^2 g(x)^2$ 
in comparison with the terms in \eq{g00-approx}
is given in figure \ref{fig:U}.
This clearly shows that the Newtonian potential dominates
for small $r$, while the vacuum energy term 
takes over for larger  $r$ until the potential is cut off 
effectively at $L_\omega$.

\paragraph{(Galactic) rotation curves.}

\begin{figure}
  \vspace{-0.2cm}
\begin{center}
 \includegraphics[scale=0.2]{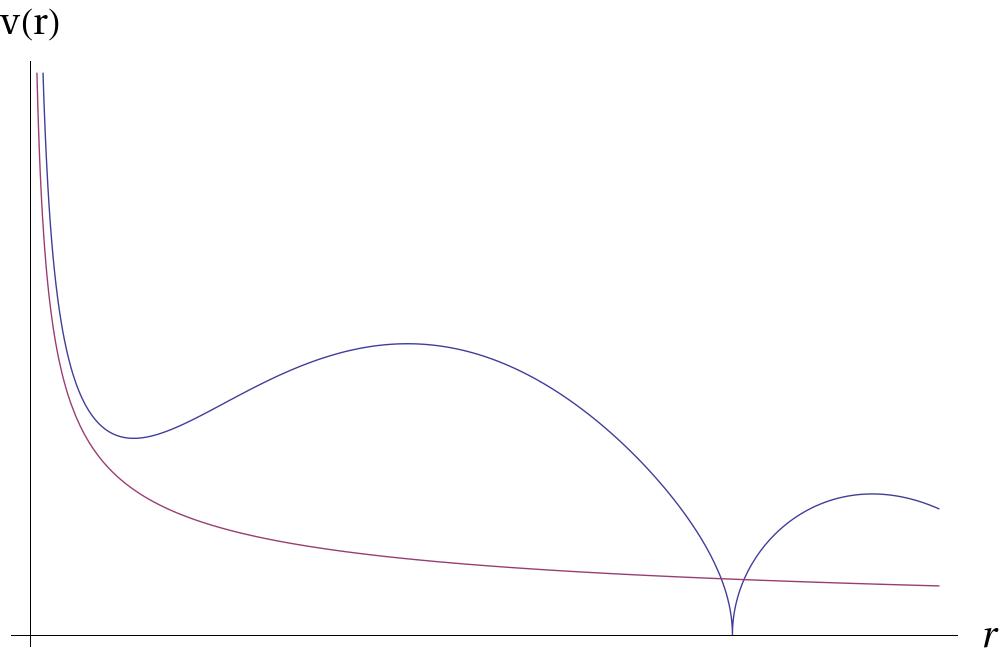}
\end{center}
\caption{Orbital velocity $v(x)$ for a central point mass
compared with Newtonian case:
Newtonian domain, enhancement and cutoff, for
$\omega = 0.1, \, g_0 = 1, \d = 0.1$.}
\label{fig:V}
\end{figure}

The non-relativistic orbital velocity $v(r)$ around a 
central mass with the above metric is given 
for small distances $r < L_\omega$ by
\be
v(r) = \sqrt{U' r} = \sqrt{2 \frac{G M}{r} 
(1+\frac{\pi^2}{3} \frac{r^2}{L_\omega^2})
 - \frac 23 \Lambda_{\rm eff} r^2} .
\label{v-SdS}
\ee
This decreases like $r^{-1/2}$ as in Newtonian
gravity as long as $E_{vac}(r) < M$,  
but for $E_{vac}(r) \approx 4\pi M$ it
starts to increase linearly
like $v \sim \sqrt{|\Lambda_{\rm eff}|}\, r$
until $r \approx L_\omega$.
At that scale, the harmonic cutoff becomes effective, 
leading to a decreasing rotational velocity
$v \sim \frac{1}{r}$
for large distances.
An illustrative plot of $v(r)$
compared with the Newtonian approximation 
is shown in figure \ref{fig:V}.
This idealized result hold only outside 
of the mass distribution, and will be modified
e.g. in galaxies. 
One naturally obtains a 
slightly increasing rotation curve, 
which is indeed often observed. 
It remains to be seen whether this allows to 
explain the galactic rotation curves,
providing an (partial?) alternative for dark matter.
We only emphasize here that
the qualitative behavior certainly goes in the right direction,
and $G = \frac{g_0^2\omega^4}{\Lambda_1^4}$  
may differ somewhat from galaxy to galaxy.

\subsection{Local linearized effective action}
\label{sec:local-bag}

The above analysis requires a non-linear treatment 
of the large-scale gravity bag. We now provide a
local, linearized description of the same 
mechanism\footnote{I am indebted to 
N. Arkani-Hamed for illuminating discussions and insights
related to this section.}. 
Consider a local neighborhood of a point $p$ inside a 
gravity bag, using ``free-falling''
normal embedding coordinates
$\partial_\mu\phi_0^i|_p = 0,\,\, g_{\mu\nu}|_p = \eta_{\mu\nu}$.  
Then the background gravity bag manifests itself in terms
of a non-vanishing 2nd derivatives of the 
embedding\footnote{$K_{\mu\nu}^i$ is essentially the 
2nd fundamental form.}
\be 
\partial_\mu\partial_\nu \phi_0^i = K_{\mu\nu}^i,
\qquad \phi_0^i(x) =  \frac 12 K_{\mu\nu}^i x^\mu x^\nu + O(x^3).
\label{Kmunu-def}
\ee
For example, a spherical harmonic gravity bag 
$\phi_0(x) = g_0\cos(\omega t)\frac{\sin(\omega r)}{\omega r}$
leads near its center 
to $\frac 12 K_{\mu\nu} x^\mu x^\nu = -\frac{1}2 g_0 \omega^2
(t^2 +\frac{1}3(x^2+y^2+z^2))$.
The advantage of this approach is that it 
allows to explore more general
backgrounds, and indeed  Ricci-flat metric perturbations 
are obtained for $K_{\mu\nu}^i \sim \d_{\mu\nu}$.
The linearized action valid near $p$ is given by
\bea
S_{lin}[\phi^i] &=& \int d^4 x \, (-\L_1^4 h_{\mu\nu} \eta^{\mu\nu} + 
 8\pi \tilde T^{\mu\nu} h_{\mu\nu})  \nn\\
 &=& \int d^4 x \,(\L_1^4 \phi^i \Box \phi^j \,\d_{ij} 
- 8\pi \phi^i\,\tilde
T^{\mu\nu} \partial_\mu\partial_\nu \phi^j\, \d_{ij}) 
\label{matter-g-lin-phi}
\eea
using partial integration, recovering \eq{eom-general-linear}.
We now add a small perturbation
corresponding to a small mass distribution $\rho$,
\be
\phi(x) = \phi_0(x) + \d \phi(x), 
\qquad\Box \phi_0 = 8\pi \frac{\tilde T^{\mu\nu}}{\L_1^4} K_{\mu\nu}.
\ee
which we assume to be static with 
$\tilde T^{\mu\nu} = \delta^{\mu,0}\delta^{\nu,0}\rho(x)$
for simplicity. We can then assume that $\d \phi(x)$
is static 
(more precisely $\partial_0^2 \d \phi \ll K_{00}$), 
so that the effective action for $\d \phi(x)$
can be written as
\bea
S_{lin}[\d\phi^i] &=& 
\int d^4 x \,(\L_1^4 \d\phi^i \Box \d\phi^j \,\d_{ij} 
- 8\pi \d\phi^i\, \tilde T^{\mu\nu} 
(2 \partial_\mu\partial_\nu \phi^j_0 +
\partial_\mu\partial_\nu \d\phi^j)\, \d_{ij} ) \nn\\
 &\approx& \int d^4 x \,(\L_1^4 \d\phi^i \Box \d\phi^j \,\d_{ij} 
- 16\pi \d\phi^i\,\rho K_{00}^j
\, \d_{ij}) 
\eea
omitting background terms.
This is  now reminiscent of scalar gravity 
where $\d \phi$ couples to $\rho$, 
cf. \cite{Shapiro:1993vr}. 
The equation of motion for $\d\phi^i$ becomes
\be
\Delta \d\phi^i = 8\pi \tilde T^{\mu\nu} K_{\mu\nu} \frac{1}{\L_1^4}
 \approx 8\pi \rho K_{00} \frac{1}{\L_1^4} 
\quad + \mbox{(gravity bag)} ,
\label{d-phi-eq}
\ee
cf. \eq{eom-embedd-2}.
The effective metric  has the form
\bea
g_{\mu\nu} &=& \eta_{\mu\nu} 
+\partial_\mu\phi_0^i\partial_\nu\phi^j_0\d_{ij}
+\partial_\mu\phi_0^i\partial_\nu\d\phi^j\d_{ij}
+\partial_\mu\d\phi^i\partial_\nu\phi_0\d_{ij}
+ \partial_\mu\d\phi^i\partial_\nu\d\phi^j \d_{ij}
\label{lin-metric-pert}
\eea
The mixed $g_{0i}$ components can be eliminated
using a change of variables
${x^\mu}' = x^\mu + \xi^\mu , \,
\xi^\mu = - \eta^{\mu\mu'} \partial_{\mu'}\phi_0^i \d\phi^j \d_{ij}$
which gives 
\bea
g_{\mu\nu}' &=&\eta_{\mu\nu}  - 2 K_{\mu\nu}^i  \d\phi^j \d_{ij} 
+  \partial_\mu\d\phi^i\partial_\nu\d\phi^j \d_{ij} .
\eea
Thus we recover the Newtonian potential 
$U(x) =  K_{00}^i \d\phi^j\d_{ij}$ as in \eq{U-newton-linfluct},
which according to \eq{d-phi-eq} satisfies
\be
\Delta U(x) = 8\pi (K_{00}^i  K_{\mu\nu}^j\d_{ij}) \tilde T^{\mu\nu}
\frac{1}{\L_1^4}
 \approx 4\pi G \rho \quad + \mbox{(gravity bag)}
\label{U-poisson-linear}
\ee
The gravitational constant $G$ is 
determined by 
\be
G = 2\frac{K_{00}^i K_{00}^j \d_{ij}}{\L_1^4},
\label{Newton-general}
\ee
in agreement with \eq{Newton-constant}.
In the case of a point mass $M$ and a harmonic gravity bag, this gives
\bea
g_{00} &=& -(1- 2 \frac{G M}{r}),  \qquad
g_{ij}  = \d_{ij}(1 + \frac 23 \frac{G M}{r})
+ O(\d g^2) .
\eea 
The factor $\frac 13$  agrees with our previous
result \eq{grr-simple-2}, and stems from the harmonic 
form of the gravity bag.
It is not hard to see \cite{blaschke-steinacker} that 
the leading Ricci-flat metric perturbation
\be
g_{\mu\nu}' =\eta_{\mu\nu} - 2  U(x) \delta_{\mu\nu} ,
\qquad U(x) = \frac 12 K \d\phi
\ee
of linearized  GR is obtained for $K_{\mu\nu} =\frac 12 \delta_{\mu\nu} K$. 
For static $\d\phi$, this gives for the Einstein tensor
$\cG_{00} \sim 8\pi G \rho$
and $\cG_{ii}^{(1)}=0$, consistent with GR.
In other words, a gravity bag with the form
\bea
\tilde\phi_0 = \cos(\omega t) \frac{\sin(k r)}{k r}
\approx (1-\frac 12 \omega^2(t^2+x^2+y^2+z^2)),  
\nn
\eea
with $k^2 = 3 \omega^2$ 
would lead to a Ricci-flat local deformation $h_{\mu\nu}$; however it is 
not clear why such a background should arise.

We conclude by pointing out that 
the effective gravitational constant $G$ in the harmonic branch
is governed by the extrinsic geometry $K_{\mu\nu}^i$ 
of a ``gravity bag'', 
whose intrinsic geometry may be (almost) flat. This is an interesting 
mechanism which deserves to be studied in detail, and
it remains to be seen whether a more sophisticated version will 
modify the detailed properties and perhaps reconcile 
it with general relativity.

\section{Conclusion}

The foundations of the matrix-model
approach to (emergent) gravity are presented. 
The effective geometry of noncommutative space-time branes 
$\cM_\theta \subset \R^D$ is identified
in the semi-classical limit, and the effective action for
fields, matter and geometry is given.  Fluctuations of the matrices 
around such a background lead to fields
and matter on $\cM_\theta$, and the commutators in the matrix model 
become derivative operators which act on these fields. 
Hence the noncommutative nature of $\cM_\theta$ is essential for this mechanism,
nevertheless $\theta^{\mu\nu}$
is largely hidden from low-energy physics and 
enters only through the effective metric.
The matrix model action can be written in a covariant i.e. 
geometric way, leading to reasonable effective 
actions which can be studied further.
In particular the $D=10$ models, notably the IKKT model,
allow to describe generic 4-dimensional 
geometries, and can be expected to define a well-behaved quantum theory.
Moreover, the Einstein-Hilbert 
action are obtained either upon quantization or as an additional
higher-order term in the matrix model.
Therefore the matrix model 
should provide a realization of Einstein gravity or some closely related 
gravity theory.
Emergent gravity from matrix models therefore becomes
a serious candidate for a realistic 
theory of gravity at the quantum level.

There are several aspects 
which make this framework very attractive from a theoretical point of view.
First, its definition requires no classical-geometrical 
notions of geometry whatsoever. The geometry arises
dynamically, which is very appealing from
the point of view of quantum gravity and cosmology. 
Another fascinating aspect is that 
the matrix model framework leads naturally to a unified picture 
of gravitons and nonabelian gauge fields, which arise as
abelian resp. nonabelian fluctuations of the basic matrices
(covariant coordinates) around a geometrical background.
The quantization around a such a background
should be technically rather straightforward, similar
to nonabelian gauge field theory. 
Nevertheless, more analytical and numerical work
on the emergence and stability of 4-dimensional NC branes in matrix
models is needed, particularly for the supersymmetric case
\cite{Nishimura:2001sx,Azeyanagi:2008bk}.

Leaving aside its theoretical appeal, 
we can briefly summarize the physical aspects of the model 
as follows. 2 types of solutions have been identified, dubbed
``harmonic branch'' and ``Einstein branch''.
The harmonic branch essentially describes branes with tension.
The most interesting feature is
that it naturally predicts a (nearly-) flat universe, resulting in
luminosity curves e.g. for type Ia supernovae which are close 
to the observed ones
(usually interpreted in terms of cosmic acceleration)
without any fine-tuning. These and other
long-distance modifications of gravity 
might also offer a mechanism for (partially?) explaining
the galactic rotation curves, without requiring large 
amounts of dark matter. Moreover, flat space 
remains to be a solution even in the preesnce of large vacuum energy, which
is very interesting in the context of the cosmological constant problem.
However, the solar system precision tests may be a problem in the
harmonic branch.
These should be obviously satisfied in 
the Einstein branch, which has not yet been studied in this framework  
in great detail; see \cite{blaschke-steinacker} for 
a realization of the Schwarzschild geometry. Generally speaking,
the physical properties of emergent gravity are not 
yet sufficiently well understood. 
This applies in particular for the would-be  $U(1)$ gauge field
 which appear to behave as additional gravitational waves.

We conclude by stressing
that the  gravity ``theory'' under consideration is not based on 
some theoretical expectations, but is simply the result of a careful 
analysis of the semi-classical limit of this type of matrix models.
Taking simplicity as a guideline towards a 
more fundamental theory, the matrix model 
appears to be an extremely appealing candidate
for a quantum theory of fundamental interactions including gravity, and 
certainly deserves a thorough investigation.

\paragraph{Acknowledgments}

I would like to thank N. Arkani-Hamed 
for illuminating discussions and hospitality at the IAS
Princeton, as well as D. Blaschke, R. Brandenberger, H. Grosse,  
P. Aschieri, D. Klammer, A. Schenkel, P. Schreivogl and P. Schupp 
for many discussions on various related topics. 
This work was supported by the FWF project P21610.

\section*{Appendix A: Some identities}

The following is a useful identity for Poisson tensors:
\bea
\partial_\mu \theta^{\mu\nu} &=& 
-\theta^{\mu\mu'}\partial_\mu \theta^{-1}_{\mu'\nu'} \theta^{\nu'\nu}  
= \theta^{\mu\mu'}\theta^{\nu'\nu} 
(\partial_{\mu'} \theta^{-1}_{\nu'\mu} 
 + \partial_{\nu'} \theta^{-1}_{\mu\mu'})  \nn\\
&=& - \theta^{\mu\mu'}\partial_{\mu'}\theta^{\nu'\nu}  \theta^{-1}_{\nu'\mu}
- \theta^{\nu'\nu} \partial_{\nu'}\theta^{\mu\mu'}\theta^{-1}_{\mu\mu'}  \nn\\
&=& - \partial_{\mu'}\theta^{\mu'\nu}  
- 2\theta^{\nu'\nu} \rho^{-1}\partial_{\nu'}\rho
\eea
where $\rho = \theta^{-n}$ is the symplectic volume, noting that 
$2\rho^{-1}\partial_{\nu}\rho = \partial_{\nu}\theta^{\mu\mu'}\theta^{-1}_{\mu\mu'}$.
This implies
\be 
\partial_\mu (\rho\,\theta^{\mu\nu}) \equiv 0  \,. 
\ee
For our restricted class of metrics,
this identity together with 
$|G_{\mu\nu}|^{1/2} = \rho e^{\sigma}$  implies
\bea
\Gamma^\mu &=& - |G_{\rho\sigma}|^{-1/2}
\partial_\nu (G^{\nu\mu}\,|G_{\rho\sigma}|^{1/2})  \nn\\
&=&  - \frac 1\rho e^{-\sigma}\,
\partial_\nu (\rho\,\theta^{\nu\nu'}\theta^{\mu\mu'}
 g_{\mu'\nu'}(x)) \nn\\ 
&=&  - e^{- \sigma}\,\theta^{\nu\nu'}
\partial_\nu (\theta^{\mu\mu'} g_{\mu'\nu'}(x)) 
\label{tilde-Gamma-vanish}
\eea

\section*{Appendix B: Identity for $\theta^{\mu\nu}$}

To obtain a covariant equation for $\theta^{\mu\nu}$,
recall the following identity derived in 
(\cite{Steinacker:2008ri}, Appendix B)
\bea
&& G^{\gamma\eta }(x)\, \nabla_\gamma \theta^{-1}_{\eta \nu}
= G^{\gamma\eta }\,\partial_\gamma \theta^{-1}_{\eta \nu} 
- G^{\gamma\eta }\, \Gamma_{\gamma\nu}^\mu \theta^{-1}_{\eta \mu}
- \Gamma^\mu \theta^{-1}_{\mu\nu} \nn\\
&&= G_{\rho\nu}\,\theta^{\rho \mu} 
\(\partial_\mu(e^{-2\sigma}\eta)
  + e^{-\sigma}\,\partial_\mu\phi\, 
G^{\gamma\eta }\partial_\gamma \partial_\eta \phi 
 - e^{-\sigma} G^{\gamma\eta} g_{\mu\eta }\,\partial_\gamma \sigma
 + 2 e^{-2\sigma}\eta\, \partial_\mu \sigma \)
- \Gamma^\mu \theta^{-1}_{\mu\nu}\, \nn
\eea
which in (any) matrix coordinates can be written as
\bea
G^{\gamma\eta }(x)\, \nabla_\gamma \theta^{-1}_{\eta \nu}
&=& G_{\rho\nu}\,\theta^{\rho \mu}
 \(e^{-2\sigma}\partial_\mu\eta
  + e^{-\sigma}\,\partial_\mu x^a\, 
G^{\gamma\eta }\partial_\gamma \partial_\eta x^a
 - e^{-\sigma}G^{\gamma\eta }  g_{\mu\eta }\,\partial_\gamma \sigma\)
-\Gamma^\mu \theta^{-1}_{\mu\nu}\, \nn\\
&=& G_{\rho\nu}\,\theta^{\rho \mu} 
\(e^{-2\sigma}\partial_\mu\eta  + e^{-\sigma}\,\partial_\mu x^a\, 
(\Box_{G} x^a + \Gamma^\eta\partial_\eta x^a)
 - e^{-\sigma} G^{\gamma\eta }  g_{\mu\eta }\,\partial_\gamma \sigma\)
- \Gamma^\mu \theta^{-1}_{\mu\nu}\, \nn\\
&=& G_{\rho\nu}\,\theta^{\rho \mu}
\(e^{-2\sigma}\partial_\mu\eta
  + e^{-\sigma}\,\partial_\mu x^a\, \Box_{G} x^a\) \nn\\
&& + (e^{-\sigma}\, G_{\rho\nu}\,\theta^{\rho \mu} g_{\mu\eta})\Gamma^\eta 
- \Gamma^\mu \theta^{-1}_{\mu\nu}\, 
-  G^{\gamma\eta } (e^{-\sigma}\,G_{\rho\nu}\, \theta^{\rho \mu}
 g_{\mu\eta })\,\partial_\gamma \sigma \nn\\
&=& G_{\rho\nu}\,\theta^{\rho \mu}
\(e^{-2\sigma}\partial_\mu \eta
  + e^{-\sigma}\,\partial_\mu x^a\, \Box_{G} x^a\) 
+ G^{\gamma\eta } \, \theta^{-1}_{\nu\eta }\,\partial_\gamma \sigma
\label{theta-covar-id}
\eea
which gives
\be
G^{\gamma\eta }(x)\, \nabla_\gamma
(e^{\sigma}\theta^{-1}_{\eta \nu})
= G_{\rho\nu}\,\theta^{\rho \mu}
\(e^{-\sigma}\partial_\mu \eta
  + \,\partial_\mu x^a\, \Box_{G} x^a\)
\label{theta-covar-id-2-app}
\ee
This is interesting, because it shows that $\theta^{\mu\nu}$
captures some of the extrinsic geometry of $\cM\subset \R^D$
through the last term.
In particular, the conservation law 
\eq{conservation-general-2} implies
\be
\partial_\mu x^a\, \Box_{G} x^a =0
\ee
\cite{Steinacker:2008ya}, which 
holds identically for $g_{\mu\nu} = G_{\mu\nu}$.

\end{document}